\documentclass{article}

\usepackage[preprint]{neurips_2023}

\usepackage[T2A,T1]{fontenc}        
\usepackage[utf8]{inputenc}     
\usepackage[russian,english]{babel}  

\usepackage{hyperref}       
\usepackage{url}            
\usepackage{booktabs}       
\usepackage{amsfonts}       
\usepackage{nicefrac}       
\usepackage{microtype}      
\usepackage{xcolor}         
\usepackage{comment}   
\usepackage{listings}
\usepackage{mdframed} 
\usepackage{graphicx}
\usepackage{fancyvrb}

\usepackage{xspace}
\usepackage{amsmath}
\usepackage{booktabs}
\usepackage{multirow}
\usepackage{array}
\usepackage{lipsum}
\usepackage{fancyvrb}
\usepackage{subcaption}
\usepackage{listings}
\usepackage{todonotes}
\usepackage{url}

\colorlet{darkblue}{blue!50!black}
\colorlet{darkred}{red!50!black}

\hypersetup{
  colorlinks = true,
  citecolor = darkblue,
  linkcolor = darkred
}

\title{Look Before You Leap: A Universal Emergent Decomposition of Retrieval Tasks in Language Models}

\author{%
  Alexandre Variengien\thanks{Work done during an internship at Conjecture. Correspondence to \texttt{alexandre.variengien@gmail.com}.} \\
  École Normale Supérieure de Lyon \\
  École Polytechnique Fédérale de Lausanne \\
  \And
  Eric Winsor \\ 
  Conjecture \\
}

\newcommand{\av}[1]{{\color{orange}\textbf{AV:} #1}}

\newcommand{\E}{\mathbf{E} }

\newcommand{\M}{\mathcal{M} }
\newcommand{\prompt}[1]{“\texttt{#1}”}

\begin{document}

\maketitle

\begin{abstract}


When solving challenging problems, language models (LMs) are able to identify relevant information from long and complicated contexts. To study how LMs solve retrieval tasks in diverse situations, we introduce ORION, a collection of structured retrieval tasks spanning six domains, from text understanding to coding. Each task in ORION can be represented abstractly by a request (e.g. a question) that retrieves an attribute (e.g. the character name) from a context (e.g. a story).
We apply causal analysis on 18 open-source language models with sizes ranging from 125 million to 70 billion parameters. We find that LMs internally decompose retrieval tasks in a modular way: middle layers at the last token position process the request, while late layers retrieve the correct entity from the context. 
After causally enforcing this decomposition, models are still able to solve the original task, preserving 70\% of the original correct token probability in 98 of the 106 studied model-task pairs. We connect our macroscopic decomposition with a microscopic description by performing a fine-grained case study of a question-answering task on
Pythia-2.8b. Building on our high-level understanding, we demonstrate a proof of concept application for scalable internal oversight of LMs to mitigate prompt-injection while requiring human supervision on only a single input. Our solution improves accuracy drastically (15.5\% $\rightarrow$ 97.5\% on Pythia-12b). This work presents evidence of a universal emergent modular processing of tasks across varied domains and models and is a pioneering effort in applying interpretability for scalable internal oversight of LMs. Code available at \url{https://github.com/aVariengien/causal-checker}.

\end{abstract}

\section{Introduction}

Recent advances in language models (LMs) \citep{vaswani2017attention} have demonstrated their flexible problem-solving abilities and their expert-level knowledge in a wide range of fields \citep{bubeck2023sparks, openai2023gpt4}. Researchers have developed a series of techniques such as fine-tuning \citep{ouyang2022training} and Reinforcement Learning from Human Feedback (RLHF) \citep{ouyang2022training} to ensure models output honest and helpful answer. However, as their abilities reach human level, supervision from human feedback becomes costly and even impossible. This necessitates more efficient or automated methods of supervision, known generally as \textit{scalable oversight}.

Moreover, existing methods only control for the output of the model while leaving the internals of the model unexamined \citep{casper2023open, ngo2023alignment}. This is a critical limitation as many internal processes can elicit the same output while using trustworthy or untrustworthy mechanisms. For instance, we would like to know whether a model answers faithfully based on available information or simply gives a users' preferred answer \citep{perez2022discovering}. We call this problem \textit{internal oversight.}


Approaches such as representation engineering \citep{zou2023repre} have demonstrated promising angles for intervening on language models at a high level. However, because such approaches do not rely on mechanisms within LM internals, they are necessarily coarse-grained in their application. On the other hand, recent works on mechanistically interpreting LMs have shown success on narrow tasks \citep{wang2022interpretability, nanda2023progress}. Some have provided insight into factual recall \citep{geva2023dissecting} and in-context learning \citep{olsson2022context}. Causal interventions have even been used to understand how models encode tasks from few shot examples \citep{hendel2023incontext} or bind entity to attributes \citep{feng2023language}. However, these works are still scoped to relatively narrow contexts and lack demonstration of concrete applications.

In this work, we study how LMs solve retrieval tasks, i.e. in-context learning problems that involve answering a request (e.g. ``What is the city of the story?'') to retrieve a keyword (e.g. ``Paris'')  from a context (e.g. a story). 

We start by introducing ORION, a collection of 15 datasets of retrieval tasks spanning six different domains from question answering to coding abilities and variable binding. We systematize the task structure by annotating each textual input with an abstract representation where the context is a table of attributes, and the request as a simple SQL-like query, e.g. ``\texttt{SELECT \textrm{\textit{Name}} FROM \textrm{Context} WHERE \textrm{\textit{Role}=City}}'' as illustrated in Figure \ref{fig:table_example_orion}.

\begin{figure}
    \centering
    \includegraphics[width=1.\textwidth]{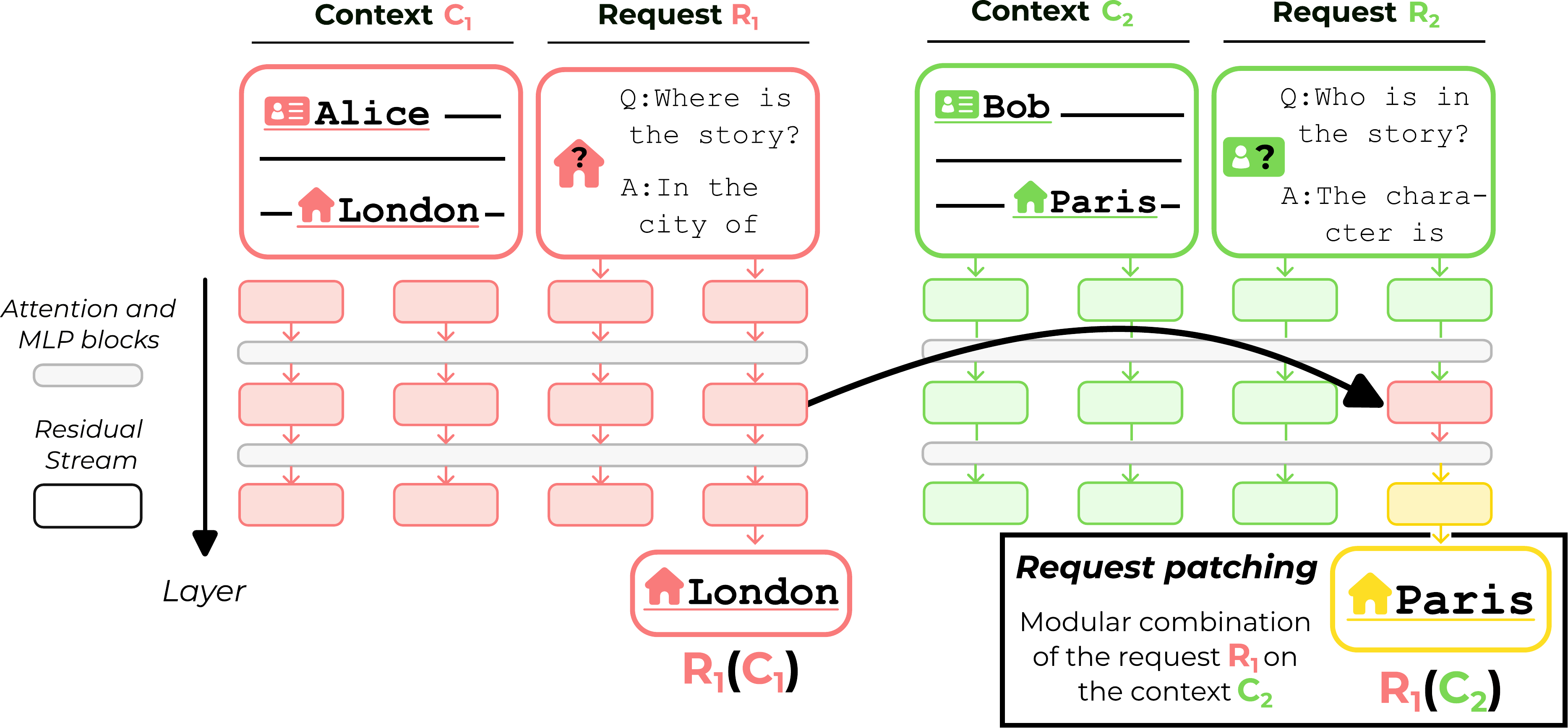}
    \caption{Illustration of our main experimental discovery. Patching the mid-layer residual stream on a retrieval task from ORION causes the language model to output a modular combination of the request from $x_1$ (asking for the city) and the context from $x_2$ (a story about Bob in Paris). We call this phenomenon \textit{request-patching}. }
    \label{fig:request_patching_key_figure}
\end{figure}

We apply causal analysis \citep{pearl2009causality, vig2020investigating, geiger2021causal} to 18 open source LMs ranging in size from 125 million to 70 billion parameters to investigate the successive role of layers at the last position on tasks from ORION. The shared abstract representation enables us to define and interpret experiments across tasks and models at scale, without the need for setting-specific labor. We discover that language models handle retrieval tasks by cleanly separating the layers at which they process the request and the context at the last token position. We design an interchange intervention \citep{geiger2021causal} called \textit{request-patching} that artificially forces this division as illustrated in Figure \ref{fig:request_patching_key_figure}. It removes access to the context for the layers representing the request and removes access to the request for the layer processing the context. Despite strongly constraining access to input information, models after this intervention preserve more than 70\% of the baseline probability of the correct answer for 98 out of the 106 pairs of models and tasks studied. These results suggest that there exists an emergent modular decomposition of tasks that applies across models and tasks. We complement this coarse-grained causal analysis with a finer-grained case study of a question-answering task on Pythia-2.8b \citep{biderman2023pythia}.

Understanding models internally is not enough to increase the operational safety of LMs. Raw knowledge needs to be translated into concrete applications. This is why we demonstrate that our understanding of how models solve retrieval tasks can be directly leveraged to mitigate the effect of prompt injection \citep{perez2022ignore} in a question answering task. Models are given inputs containing distractor sequences that trigger models to output a token unrelated to the task. The goal is to detect and eliminate the influence of distractors while requiring minimal human supervision. We present a proof-of-concept based on request-patching that only requires humans to verify the model output on a \textit{single} trusted input. Our technique significantly improves the performance of models on sequences with distractors (0\% $\rightarrow$ 70.5\% accuracy for Pythia-410m, 15.5\% $\rightarrow$ 97.5\% for Pythia-12b), while causing no change in performance on control inputs. To our knowledge, this is the first demonstration that scalable internal oversight of LMs is feasible.

In summary, our main contributions are as follow:
\begin{enumerate}
    \item We introduce ORION, a collection of structured retrieval tasks. It is a data-centric approach to unlocking scalable causal analysis of LMs, enabling the \textbf{comparative study of 18 models on 6 task domains.} 
    \item We discover a macroscopic \textbf{modular decomposition of retrieval tasks in LMs' internals that is universal across task domains and models.}
    \item We \textbf{link macroscopic and microscopic descriptions of LMs' internals} with a fine-grained case study of a question-answering task on Pythia-2.8b.
    \item We apply this knowledge to a \textbf{proof of concept for \textit{scalable internal} oversight of LMs} solving a retrieval task in the presence of prompt injection, requiring human supervision on only a single input.
\end{enumerate}

\section{Background} \label{sec:background}

\subsection{The Transformer architecture for autoregressive language models} 

An autoregressive language model, $\M_{\theta}$ with parameters $\theta$, maps a sequence of input tokens $x = (x_1, x_2, ..., x_n)$ to a probability distribution over the next token $x_{n+1}$. For the Transformer architecture \citep{vaswani2017attention}, we have:
\begin{align*}
    p(x_{n+1} | x) &= \M_{\theta}(x) \\
    &= \mathrm{softmax}(\pi_n(x))
\end{align*}

The pre-softmax values $\pi_n$ are the logits at the $n$-th token position. For the GPT-2 Transformer architecture \citep{gpt2} with $L$ layers, the function $\M_{\theta}$ can be further broken down as follows:

\begin{align*}
    \pi_n &= \textrm{LN}(z^L_n) W_U \\
    z^l_k &= z^{l-1}_k + a^l_k + m^l_k \\ 
    m^l_k &= \textrm{MLP}(z^{l-1}_k + a^l_k) \\
    &= \textrm{LN}\left(W_{out} \left(\textrm{GELU}(W_{in}(z^{l-1}_k + a^l_k) + b_{in})\right) + b_{out}\right) \\
    a^l_k &= \textrm{Attn}(z^{l-1}_{\leq k}) \\
    z^0_k &= W_E t + W_P
\end{align*}

The final logits $\pi_l$ are constructed by iteratively building a series of intermediate activations $z^l_k$ we call the \textit{residual stream}, following \cite{elhage2021mathematical}. The residual stream $z^l_k$ at token position $k$ and layer $l$ is computed from the residual stream at previous token positions at the previous layer $z^{l-1}_{\leq k}$ by adding the results of $\textrm{Attn}$, a multi-headed attention module, and $\textrm{MLP}$, a two-layer perceptron module. The $\textrm{MLP}$ module depends on the residual stream $z^{l-1}_k$ at position $k$ and layer $l-1$ while the attention module can aggregate information from the previous layer across every previous token position. The residual stream is initialized with the embeddings computed from the token and positional embedding matrices $W_E$, $W_P$, and the one-hot encoding of the input tokens $t$. Finally, $W_U$ is the unembedding matrix, $\textrm{GELU}$ the Gaussian error linear unit activation function \citep{hendrycks2016gaussian}, and $\textrm{LN}$ is a layer normalization function \citep{ba2016layer} applied to the final residual stream and the output of each module.

In practice, the models we study have slight deviations from the GPT-2 architecture. The Pythia \citep{biderman2023pythia}, Falcon \citep{falcon40b} and Llama 2 \citep{touvron2023llama} models use parallelized attention and MLP. In the formulae above, this translates as $m^l_k = \textrm{MLP}(z^{l-1}_k)$. Moreover, Falcon contains additional layer normalization at the input of modules. LLama 2 uses the SwiGLU activation function \citep{shazeer2020glu} and layer normalization only at the input of modules.

\subsection{Computational graph as causal graph} \label{sec:intervention_notations}

The experimental paradigm of causal analysis applied to machine learning models initiated by \citep{vig2020investigating} and \citep{geiger2021causal} treats the computational graph of a neural network as a causal graph. The goal of causal analysis is to answer questions about \textit{why} a model outputs a given answer. This requires uncovering the causal links tying the inputs to the output, as well as characterizing the role of the internal components critical for the model's function. To this end, researchers rely on \textit{causal interventions} \citep{pearl2009causality}, experiments that replace a set of activations with fixed values.

In this work, we use single-input \textit{interchange intervention}\footnote{It is sometimes called ``activation patching'' in the literature see e.g. \cite{wang2022interpretability}} \citep{geiger2021causal}. It is a simple form of causal intervention where we intervene on one variable at a time by fixing its value to be the value of that same variable on another input. We write $\M(x|A\leftarrow A(x'))$ the output of the model after the single-input interchange intervention on the \textit{target input} $x$, replacing the activation of the node $A$ by its value on the \textit{source input} $x'$. Figure \ref{fig:orion_abstract_diagram} illustrates the experiment $\M(x|z_1^1\leftarrow z_1^1(x'))$.

\section{ORION: a collection of structured retrieval tasks}


Our study concentrates on retrieval, a fundamental aspect of in-context learning, which involves answering a request (e.g. ``What is the name of the city?'') by identifying the correct attribute (e.g. a city name) from the context (e.g. a story). To facilitate this study, we crafted a collection of datasets dubbed the \textbf{O}rganized \textbf{R}etr\textbf{I}eval \textbf{O}perations for \textbf{N}eural networks (ORION).

Creating datasets for LM causal analysis requires precise control over individual tokens. However, many datasets from the existing literature are only given brief descriptions, solving the technicalities of token-level control in an ad-hoc manner. The underlying dataset design principles are seldom explicitly discussed. We address this gap during the design of ORION, by explicitly stating desiderata for the datasets and enforcing them through automatic testing. This additional structure allows for systematized, scalable causal analyses across task domains and models.

\begin{figure}
    \centering
    \includegraphics[width=1.\textwidth]{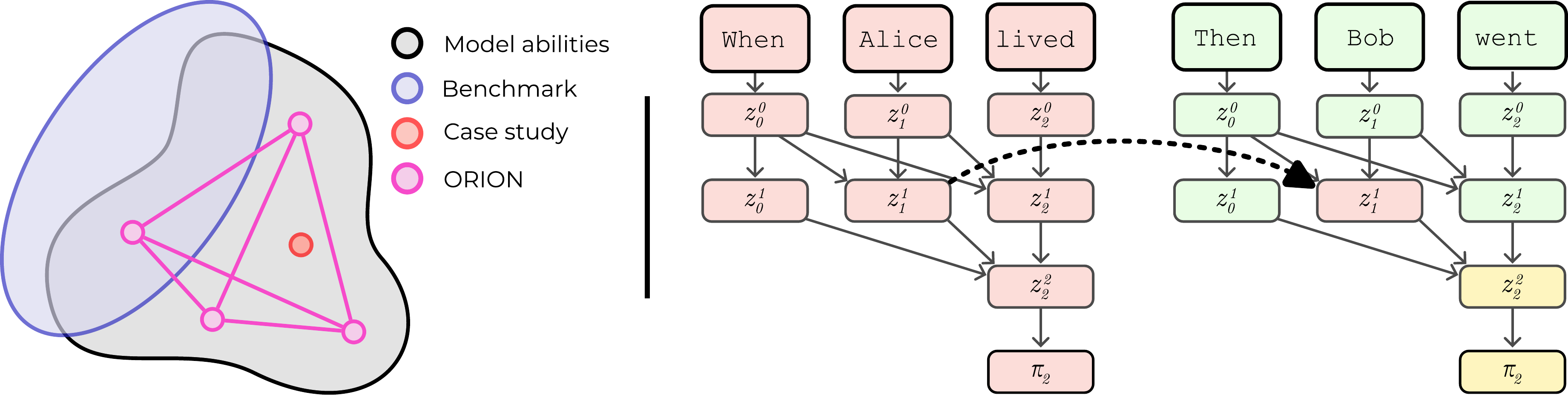}
    \caption{ Illustration of our two central methods: precise dataset design and causal intervention.
    \textbf{Left:} Informal illustration of different types of datasets for investigating the abilities of language models. Benchmarks such as MMLU aim at faithfully mapping the frontier of language models' abilities through behavioral analysis. However, because of experimental constraints, interpretability case studies are often focused on a narrow task. By introducing ORION we propose a middle-ground: it is a collection of narrow tasks spanning different domains while being connected by a common structure, enabling the design of causal experiments at scale. \textbf{Right:} An interchange intervention (here $\M(x|z_1^1\leftarrow z_1^1(x'))$) is a causal intervention where an activation (here the residual stream $z_1^1$) is replaced by its value on a source input ($x'=$``\texttt{When Alice lived}'') while the rest of the model is run on the target input ($x=$``\texttt{Then Bob went}''). The activations downstream of the intervened node depend on the \textit{two} inputs (in yellow). }
    \label{fig:orion_abstract_diagram}
\end{figure}

As illustrated informally in Figure \ref{fig:orion_abstract_diagram}, the need for granular dataset manipulations was solved in previous work by introducing strong data constraints, hence limiting their scope to narrow case studies. ORION does not aim at expanding the breadth of single analyses, but is instead a tool to run and interpret experiments on a \textit{collection} of case studies in parallel with little additional overhead. See Table \ref{tab:orion_tasks} for a succinct overview of the tasks contained in ORION.

\subsection{Abstract representation}

Each textual input (i.e. LM prompt) from ORION is annotated with an abstract representation $(C, R)$ where $C$ represents the context and $R$ the request. In the example of Figure \ref{fig:table_example_orion}, the context is a story introducing a place, a character, and an action, while the request is a question written in English asking for the city of the story. 

The context $C$ is abstractly represented as a table where each line is a list of attributes. The request $R$ is retrieving a target attribute $ATTR_t$ (e.g. the ``name'' attribute in Figure \ref{fig:table_example_orion}), from lines where a filter attribute $ATTR_f$ (e.g. the narrative role) has the value $v_f$ (e.g. ``city''). The request can be written using a language in the style of SQL as follows: \texttt{SELECT $ATTR_t$ FROM $C$ WHERE $ATTR_f = v_f$} (e.g. \texttt{SELECT \textrm{\textit{Name}} FROM \textrm{Context} WHERE \textrm{\textit{Role}=City}}). 

We note $R(C)$ the results of applying the request on the context. This is the ground truth completion for LMs evaluated on the retrieval task. In practice, $R(C)$ is a single token called the \textit{label token}. On the example we have $R(C) = $ ``\texttt{ Valencia}''.

\begin{figure}
    \centering
    \begin{minipage}{0.15\textwidth}
    \textbf{Textual input:} 
    \end{minipage}
    \hspace{3mm}
    \begin{minipage}{0.8\textwidth}
        \begin{mdframed}[align=center]
\texttt{Story: In the lively city of Valencia, a skilled veterinarian [...].
"I'm Christopher" he replied, [...]. \\ \\
Question: What is the city of the story? The story takes place in}
        \end{mdframed}
    \end{minipage}
    \\
    \vspace{2mm}
     \begin{minipage}{0.15\textwidth}
    \textbf{Abstract representation:} 
    \end{minipage}
    \hspace{3mm}
    \begin{minipage}{0.4\textwidth}
        \centering
        \textbf{Context} $C$\\
        \vspace{2mm}
        \begin{tabular}{|c|c|}
            \hline
            \textit{Name} & \textit{Role}  \\
            \hline
            \texttt{\_Valencia} & City  \\
            \texttt{\_Christopher} & Main Character  \\
            \texttt{\_veterinarian} & Character Job  \\
            \hline
        \end{tabular}
    \end{minipage}
    \hfill
    \begin{minipage}{0.4\textwidth}
        \centering
        \textbf{Request} $R$
        \begin{mdframed}[ align=center]
             \texttt{SELECT \textrm{\textit{Name}} FROM \textrm{Context} WHERE \textrm{\textit{Role}=City} }
        \end{mdframed}
    \end{minipage}

    \caption{Example input from ORION for the question-answering task. Textual inputs are annotated with an abstract representation of the context and the request. Abstract context representations are tables where each line lists attributes relative to a story element. Requests can be formulated using simple SQL-like queries. }
    \label{fig:table_example_orion}
\end{figure}

\subsection{Desiderata for datasets}

To facilitate the application of causal analysis, we enforce a list of desiderata on datasets from ORION. We explicitly define them below and share the motivation behind their definition.

\begin{enumerate}
    \item \textbf{Structured.} Every textual input in ORION accepts an abstract representation using the context and request representation defined above.
    \textit{Motivation: Providing a unified structure to define and interpret causal interventions without the need for setting-specific labor.}
    \item \textbf{Decomposable.} For every dataset $D$ in ORION, for every abstract representations $(C_1, R_1), (C_2, R_2)$ in $D$, $R_2(C_1)$ and $R_1(C_2)$ are well-defined. This means that an arbitrary request can be applied to an arbitrary context from the same task. Abstract representations of requests and contexts can be freely interchanged across a task.
    \textit{Motivation: Enabling the design of interchange interventions.}
    \item \textbf{Single token.}. For every dataset $D$ in ORION, for every abstract representations $(C_1, R_1), (C_2, R_2)$ in $D$, $R_1=R_2 \Leftrightarrow R_1(C_1)=R_2(C_1)$. In other words, in a given context, the output of each request gives a unique answer. It ensures that measuring the next-token prediction is enough to know which request has been answered.
    \textit{Motivation: Making experiments easy to measure and computationally efficient.}
    \item \textbf{Monotasking.} For every dataset $D$ in ORION, for every abstract representation $(C, R)$ in $D$, there is a unique line in $C$ such that $ATTR_f = v_f$. This condition ensures that requests are answerable with unambiguous answers.
    \textit{Motivation: Making analysis tractable. It is easier to understand models solving a single problem than solving multiple problems in parallel.}
    \item \textbf{Flexible.} ORION contains diverse tasks spanning multiple domains and levels of complexity. In practice, we demonstrate the flexibility of the ORION structure by creating 15 different tasks spanning six different language model abilities.
    \textit{Motivation: Enabling rich comparative analysis across models and domains.}
\end{enumerate}

In the code implementation, we designed automatic tests to ensure that conditions ``Decomposable'', ``Single token'', and ``Monotasking'' are respected for every task in ORION. 

\subsection{Dataset creation}

\begin{figure}
    \centering
    \includegraphics[width=0.7\textwidth]{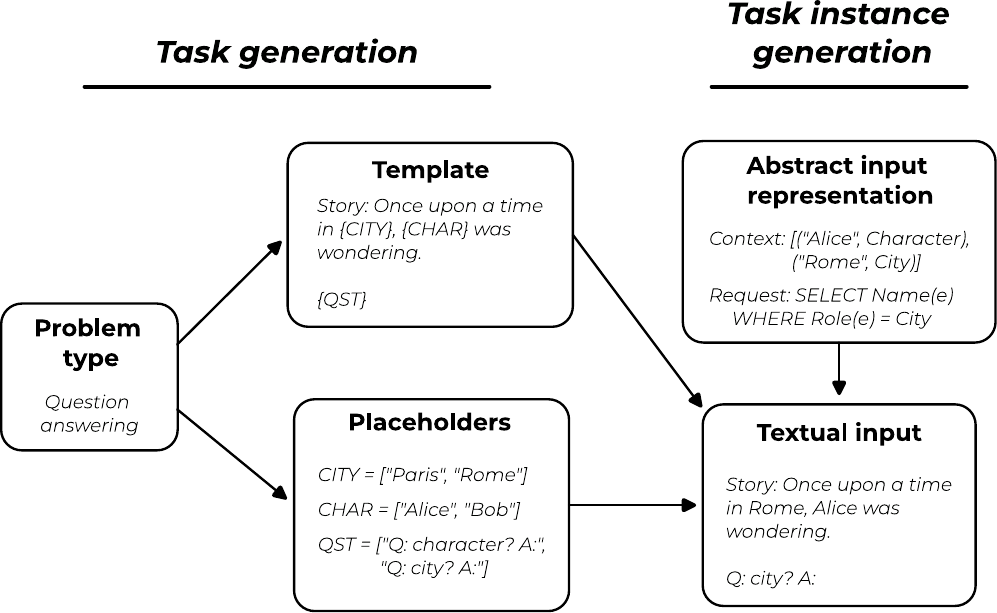}
    \caption{The semi-automatic task generation process used to create ORION. We use ChatGPT to create a template and values for the placeholders given a problem type. To generate an instance from the task, we start by randomly selecting placeholder values to create an abstract input representation. Then, we use a format string to fill the template. When we need more flexibility, we use GPT-4 to incorporate the placeholder values into the template.}
    \label{fig:task_generation}
\end{figure}

To create the ORION task datasets, we use a semi-automatic process illustrated in Figure \ref{fig:task_generation}, leveraging the creative writing ability of ChatGPT\footnote{\url{https://chat.openai.com/}}. Concretely, we use the following workflow:

\begin{enumerate}
    \item Find a problem that can be formulated as a retrieval task, e.g. question-answering.
    \item Use ChatGPT to create a template with placeholders, e.g. a story with placeholders for narrative variables such as the city, the name of the character, and the question being asked.
    \item Use ChatGPT to create a set of placeholders.
    \item Procedurally generate a set of abstract representations for the contexts and requests.
    \item Generate the textual inputs from the abstract representation using format strings or ChatGPT when more flexibility is required.
\end{enumerate}

We applied this workflow to create 15 tasks spanning six problem domains requiring different abilities: question answering, translation, factual recall, variable binding, abstract pattern matching (induction pattern) and coding. For each domain, we created variation of surface-level parameters of the task (e.g. changing language of the translation). We give an example input-output pair for each in Table \ref{tab:orion_tasks}. We provide a detailed discussion about task choices as well as a precise description of each dataset in Appendix \ref{app:orion}. 

\begin{table}[ht]
  \centering
    \begin{tabular}{p{0.16\textwidth}p{0.52\textwidth}p{0.15\textwidth}p{0.1\textwidth}}
    \hline
    \textbf{Task name} & \textbf{Example Prompt} & \textbf{Label token} & \textbf{Variations} \\
    \hline
    Question-answering \newline(base) 
    &
        \begin{minipage}[t]{0.4\textwidth}
            \begin{Verbatim}[fontsize=\scriptsize, commandchars=\\\{\}]
Story: In the lively city of Valencia, spring mornings 
\textrm{[...]} as a skilled veterinarian \textrm{[...]} "I'm Christopher" 
he replied, \textrm{[...]}. 

Question: What is the name of the main character?
Answer: The main character is named
            \end{Verbatim}
            \end{minipage}
            \vspace{1mm}
    &
    \texttt{\_Christopher} & 1 \\ \midrule
    Question-answering \newline(uniform prefix) 
    &
        \begin{minipage}[t]{0.4\textwidth}
            \begin{Verbatim}[fontsize=\scriptsize, commandchars=\\\{\}]
Story: In the lively city of Valencia, spring mornings 
\textrm{[...]} as a skilled veterinarian \textrm{[...]} "I'm Christopher" 
he replied, \textrm{[...]}.

Question: What is the name of the main character?
Answer: The answer is "
            \end{Verbatim}
            \end{minipage}
            \vspace{1mm}
    &
    \texttt{Christopher} & 1 \\ \midrule
    Question-answering \newline(question first) 
    &
        \begin{minipage}[t]{0.4\textwidth}
            \begin{Verbatim}[fontsize=\scriptsize, commandchars=\\\{\}]
Question: What is the name of the main character?

Story: In the lively city of Valencia, spring mornings 
\textrm{[...]} as a skilled veterinarian \textrm{[...]} "I'm Christopher"
he replied, \textrm{[...]}.

Answer: The answer is "
            \end{Verbatim}
            \end{minipage}
            \vspace{1mm}
    &
    \texttt{Christopher} & 1 \\ \midrule
    Question-answering \newline(mixed \newline templates) 
    &
        Uniform distribution of prompts from three variations of question-answering above.
    &
    \texttt{ } & 1 \\ \midrule
    Translation
    &
        \begin{minipage}[t]{0.4\textwidth}
            \begin{Verbatim}[fontsize=\scriptsize, commandchars=\\\{\}]
English:
In an era defined by increasing global temperatures \textrm{[...]}
At the forefront is M. Smith, a marine biologist \textrm{[...]}
Next, we turn to M. Miller, a climate economist \textrm{[...]}

French:
[...]
Nous nous tournons ensuite vers M.
            \end{Verbatim}
            \end{minipage}
            \vspace{1mm}
    &
    \texttt{\_Miller} & 3 \\ \midrule 
    Factual recall
    &
        \begin{minipage}[t]{0.4\textwidth}
            \begin{Verbatim}[fontsize=\scriptsize, commandchars=\\\{\}]
Question: On which continent did Muhammad Ali live?
Answer:
            \end{Verbatim}
            \end{minipage}
            \vspace{1mm}
    &
    \texttt{\_America} & 2 \\ \midrule 
    Variable \newline binding
    &
        \begin{minipage}[t]{0.4\textwidth}
            \begin{Verbatim}[fontsize=\scriptsize, commandchars=\\\{\}]
Anthony has a collection of pencils. 50 pencils are 
blue, 10 pencils are red, and 20 pencils are green.

How many pencils in total are either blue or green?
We'll add the number of green pencils (
            \end{Verbatim}
            \end{minipage}
            \vspace{1mm}
    &
    \texttt{20)\_} & 3 \\ \midrule 
    Induction pattern-matching 
    &
        \begin{minipage}[t]{0.4\textwidth}
            \begin{Verbatim}[fontsize=\scriptsize, commandchars=\\\{\}]
xnGWu:nJIbF
etmNX:TzgIS
ZvcIf:Gcqvs
\textrm{[...]}
AjvlA:pXMgi
etmNX:
            \end{Verbatim}
            \end{minipage}
            \vspace{1mm}
    &
    \texttt{T} & 1 \\ \midrule 
    Type hint \newline understanding
    &
        \begin{minipage}[t]{0.4\textwidth}
            \begin{Verbatim}[fontsize=\scriptsize, commandchars=\\\{\}]
def calculate_circumference(circle: Circle) -> float:
\textrm{[...]}
W = Rectangle(Point(2, 3), Point(6, 5))
D = Circle(Point(0, 0), 5)
print(calculate_circumference(
            \end{Verbatim}
            \end{minipage}
            \vspace{1mm}
    &
    \texttt{D} & 3 \\ \midrule 
    \bottomrule
\end{tabular}
\vspace{3pt} 
\caption{Tasks from the ORION collection contain varied problem type and prompt format. For readability, we use ``\textrm{[...]}'' to shorten the prompts. The rest of the text is part of the textual input. In particular, ``\texttt{[...]}'' is part of the prompt for the translation task. We use ``\texttt{\_}'' to indicate a space in the label token.
   }
\label{tab:orion_tasks}
\end{table}

\subsection{Performance metrics}
We define a task $T$ as a set of input-output pairs $(x,y)$ where $x$ is the LM input and $y$ is the expected label token. We use two main metrics to quantify the performance of a language model on an ORION task $T$.
\begin{itemize}
    \item \textbf{Accuracy:} ~ $\E_{(x,y) \sim T}[\M(x)=y]$
    \item \textbf{Token probability}: ~ $\E_{(x,y) \sim T}[p(y|x)]$
\end{itemize}

Accuracy serves as our primary metric to assess model performance in solving tasks due to its straightforward interpretation and practical application in language models, where the most probable token is often chosen. 

However, accuracy falls short in capturing nuanced aspects of predictions, for instance accuracy doesn't measure the margin by which a token is the most probable. To have a granular evaluation of model behavior after interventions, we employ token probability, offering a continuous measure. 

Finally, to remove the distorting effect of the final softmax, we used logit difference as a third secondary metric (see Appendix \ref{app:additionnal_perf_metric}).

\subsection{Model performances}

We evaluate the performance of 18 models from four different model families: GPT-2 \citep{gpt2}, Pythia \citep{biderman2023pythia}, Falcon \citep{falcon40b} and Llama 2 \citep{touvron2023llama}. We study base language models for all families except Falcon where we include two instruction fine-tuned models. We choose the models to capture diverse scale, architecture, and training techniques. 

\begin{figure}
    \centering
    \includegraphics[width=0.75\textwidth]{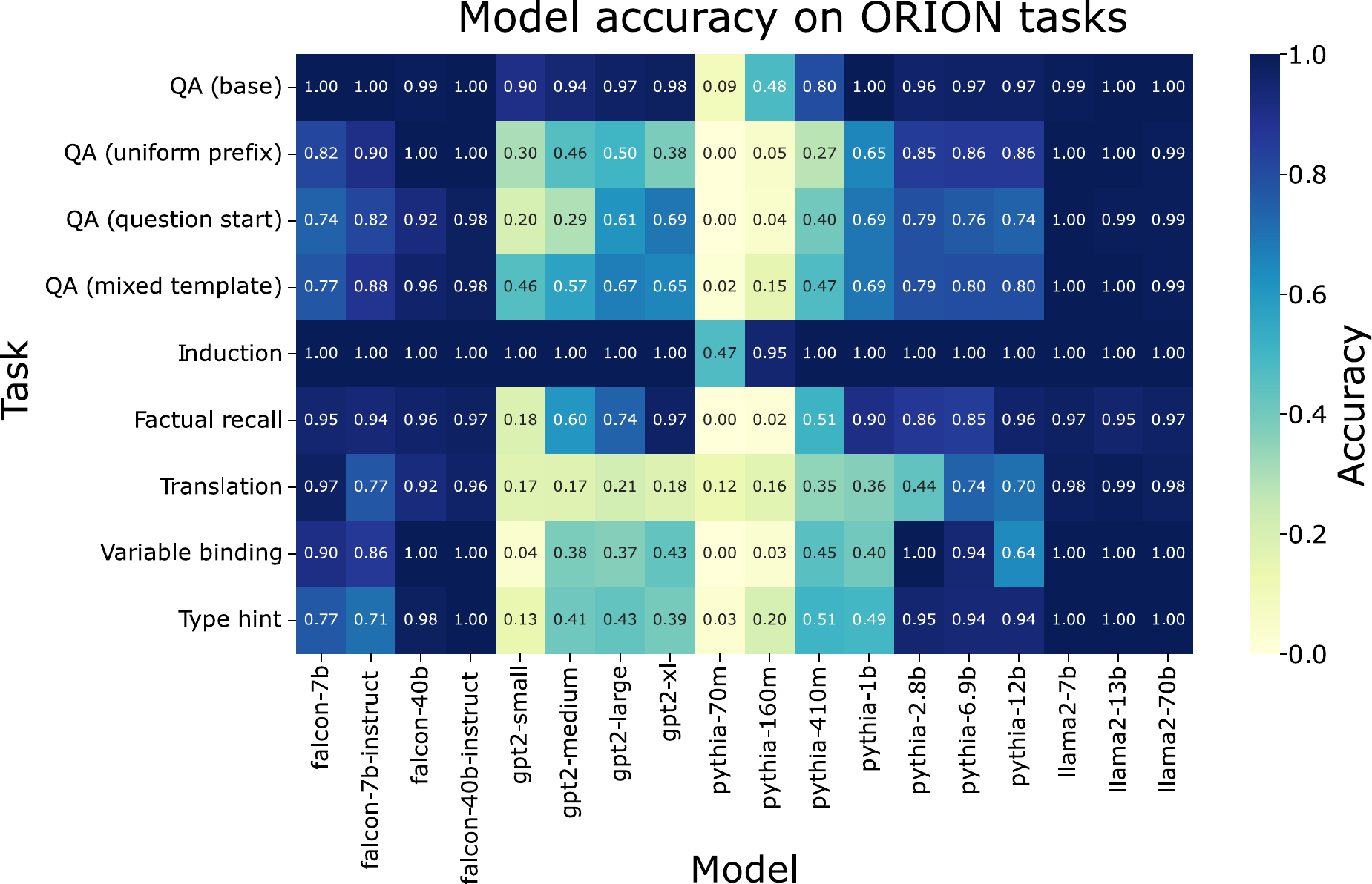}
    \caption{Accuracy of 18 models on the ORION task collection. Models with more than 7 billion parameters are able to robustly solve every task. However, simple tasks such as the base question-answering can be solved by models as small as GPT-2 small (125 million parameters), enabling comparative studies across a wide range of model scales. }
    \label{fig:baseline_perf_matrix}
\end{figure}

Figure \ref{fig:baseline_perf_matrix} shows the accuracy of every model on all tasks in ORION. Unsurprisingly, larger models can solve a wider range of problems. Nonetheless, even GPT-2 small with 125M parameters, one of the smallest models, can solve the simplest version of the question-answering task with 100\% accuracy. Evaluations using the token probability and logit difference are available in Appendix \ref{app:additionnal_perf_metric}.

In the following analyses, we only consider settings where the model can robustly solve the task. Thus, we focus on pairs of models and tasks that have greater than 70\% accuracy. 

\section{Macroscopic causal analysis on ORION: a universal emergent decomposition of retrieval tasks} \label{sec:macro_analysis}


To correctly solve retrieval tasks, an LM has to gather and combine at the last token position information coming from the request and the context. We focus our investigations on understanding how these two processing steps are organized in the intermediate layers of the last token position. 
        
In this section, we choose to consider a coarse-grained division of the model, intervening on full layers instead of a finer-grained division, e.g. considering single-attention heads and MLP blocks. We find this level of analysis is sufficient to develop a high-level causal understanding of how language models solve retrieval tasks while providing a computationally tractable set of experiments to run at scale. We complement this general coarse-grained analysis in Section \ref{sec:micro_analysis} with a finer-grained case study on Pythia-2.8b solving a question-answering task.

\subsection{Methods}

Our main experimental technique is \textit{residual stream patching}. Residual stream patching is a single-input interchange intervention, replacing the residual stream at a layer $L$ at the last position in the forward pass of the model on input $x_2$ with its activation from another input $x_1$. Following the notation introduced in Section \ref{sec:intervention_notations}, we note $\M(x_2| z^L_n\leftarrow z^L_n(x_1))$ the model output on $x_2$ after this intervention. 

As shown in Figure \ref{fig:request_patching_key_figure}, residual stream patching makes every component before layer $L$ have the activation it takes on $x_1$ while the components after layer $L$ receive mixed activations (denoted by the yellow color in the figure). These later layers see activations at the last position coming from $x_1$ while activations from earlier positions come from $x_2$.

To characterize the output of the patched model, we measure the token probability and accuracy for three different label tokens related to the inputs $x_1$ and $x_2$. We use both label tokens from the input $x_1$ and $x_2$, $R_1(C_1)$ and $R_2(C_2)$ respectively, and the label token $R_1(C_2)$ that is the result of applying the request from $x_1$ on the context of $x_2$.

To facilitate comparisons between different tasks and models, we normalize the token probability based on the mean probability of the correct token given by the model for the task. In addition, we calculate the normalized accuracy where 0 represents the accuracy of random guess, i.e. responding to a random request in a given context while 1 denotes the model's baseline accuracy for that task. It is worth noting that the normalized accuracy can fall below zero if the model's performance is poorer than randomly guessing. Normalized accuracy and token probability can also become greater than 1 if the intervention puts the model in a better situation to solve the task. In the majority of cases, causal interventions disrupt the model mechanism and thus rarely increase performance. 

We perform residual stream patching at the last position for every layer, model, and task of ORION. For each task, we use a dataset of 100 prompts and average the results of 100 residual stream patching experiments with $x_1$ and $x_2$ chosen uniformly from the task dataset. 

In the following section, we present the phenomenon of request-patching on the question-answering task on a single model, we then extend the findings to several models and then to every model and task in ORION. 

\subsection{Results of residual stream patching} \label{sec:resid_patching_results}

\begin{figure}
    \centering
    \includegraphics[width=\textwidth]{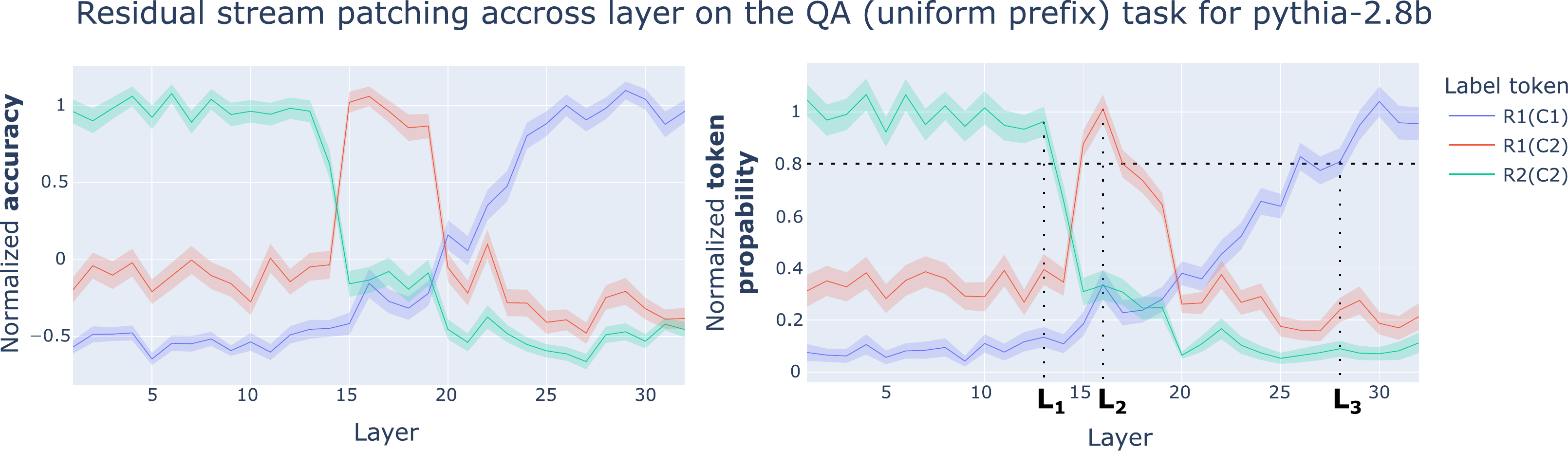}
    \caption{Normalized token probability and accuracy for the label tokens $R_1(C_1)$, $R_1(C_2)$ and $R_2(C_2)$ after patching the residual stream across all layers. Patching early (before $L_1=13$) and late (after $L_3=27$) leads to the expected results, respectively no change in output and patching the output from $x_1$. However, intervening on the middle layer ($L_2=16$) leads to the model confidently outputting the token $R_1(C_2)$, a modular combination of the request from $x_1$ and the context from $x_2$. }
    \label{fig:pythia_2_8b_3_labels}
\end{figure}

Figure \ref{fig:pythia_2_8b_3_labels} shows the results of residual stream patching on the question-answering task with a uniform answer prefix (see Table \ref{tab:orion_tasks} for an example) for the Pythia-2.8b model. We observe that after residual stream patching on the layer before layer 13, the model is outputting $R_2(C_2)$ with 100\% normalized token probability. Our interpretation is that this intervention does not perturb the model processing of $x_2$. We further observe that residual stream patching after layer 27 causes the model to output $R_1(C_1)$ with more than 80\% normalized token probability. In effect, patching the residual stream after a certain layer is equivalent to hard-coding the model output on $x_1$.

Surprisingly, when patching between layers 15 and 16, we observe that the model outputs $R_1(C_2)$ with 100\% normalized accuracy, i.e. with the same accuracy level as the baseline task accuracy. The model is outputting the results of the request contained in the input $x_1$ in the context of the input $x_2$. Moreover, Figure \ref{fig:pythia_2_8b_3_labels} shows that the more granular measure of probability of the correct token is of the same order as the baseline probability of the correct token on the task. We call this phenomenon, \textit{request-patching}, i.e. a residual stream patching experiment that leads to the $R_1(C_2)$ label token being confidently outputted by the patched model. Such results demonstrate that the causal intervention coherently intervenes in the model’s internal computation, causing it to modularly combine high-level information from two different prompts.

We observe a sudden jump in the normalized accuracy of request-patching from 0 to 1 between layers 14 and 15. However, it is likely that transforming the sequence of tokens representing the question into a representation of the request takes several layers. Thus, we hypothesize that a large part of the request processing happens at the previous token positions of the question. In this interpretation, the observed jump at layer 15 results from the intermediate representation of the request being propagated to the last position through attention modules.

\paragraph{Defining limit layers.}

From the results of the residual stream experiments, we define three layers -- $L_1$, $L_2$, and $L_3$ -- delimiting the three different outcomes of residual stream patching as shown in Figure \ref{fig:pythia_2_8b_3_labels}.

\begin{itemize}
    \item $L_1$ is the maximal layer at which the normalized token probability of the label token $R_1(C_1)$ is greater than 80\%. It marks the end of the region where residual stream patching does not interfere with the model output.
    \item $L_2$ is the layer where the normalized probability of the label $R_1(C_2)$ is maximal. It is the place where the effect of request-patching is the strongest. 
    \item $L_3$ is the minimal layer where the normalized probability of the label $R_2(C_2)$ is greater than 80\%. It marks the start of the region where residual stream patching leads to a complete overwrite of the model output.
\end{itemize}

We choose the token probability as a continuous metric to measure the model prediction. The 80\% threshold has been chosen arbitrarily as a criterion to consider that the model is mainly outputting a single label token.

\begin{figure}
    \centering
    \includegraphics[width=0.95\textwidth]{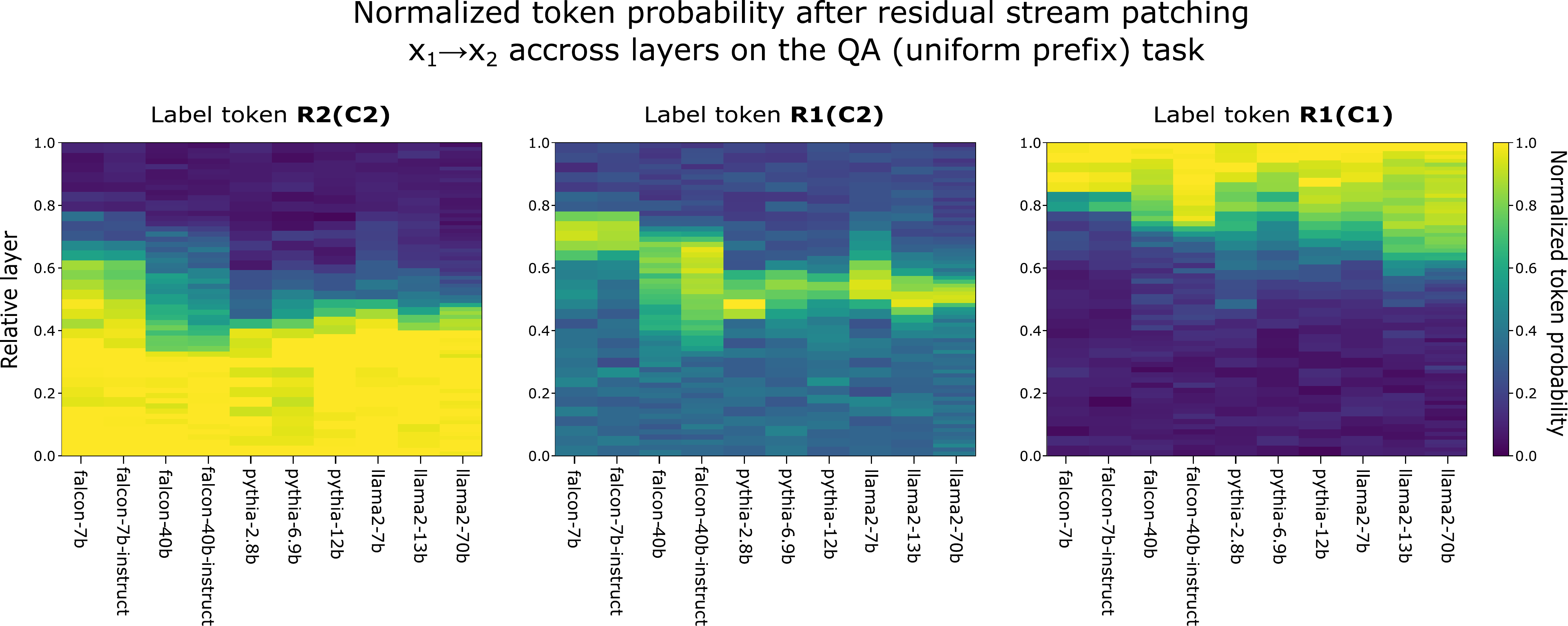}
    \caption{Normalized probability of the label tokens after residual stream patching across all layers on the question-answering task with uniform prefix. To enable comparison across models, we use the relative layer with 0 as the first and 1 as the last layer. Request-patching is general across models: mid-layer residual stream patching causes the model to output $R_1(C_2)$ with more than 80\% normalized token probability.}
    \label{fig:layer_heatmap}
\end{figure}

\paragraph{Request-patching is general across models.}
The residual stream patching results for all models able to solve the question-answering task with uniform prefixes are shown in Figure \ref{fig:layer_heatmap}. We observe a similar layer organization as the one presented for Pythia-2.8b. For all models, there exists a span of intermediate layers (40-80\% of the model depth) where residual stream patching leads the model to output $R_1(C_2)$  with a high probability (>80\% normalized probability). This span of layers seems to be the same for the base and fine-tuned Falcon models. This is coherent with the intuition that fine-tuning is only superficially affecting the model internals.

\begin{figure}
    \centering
    \includegraphics[width=0.8\textwidth]{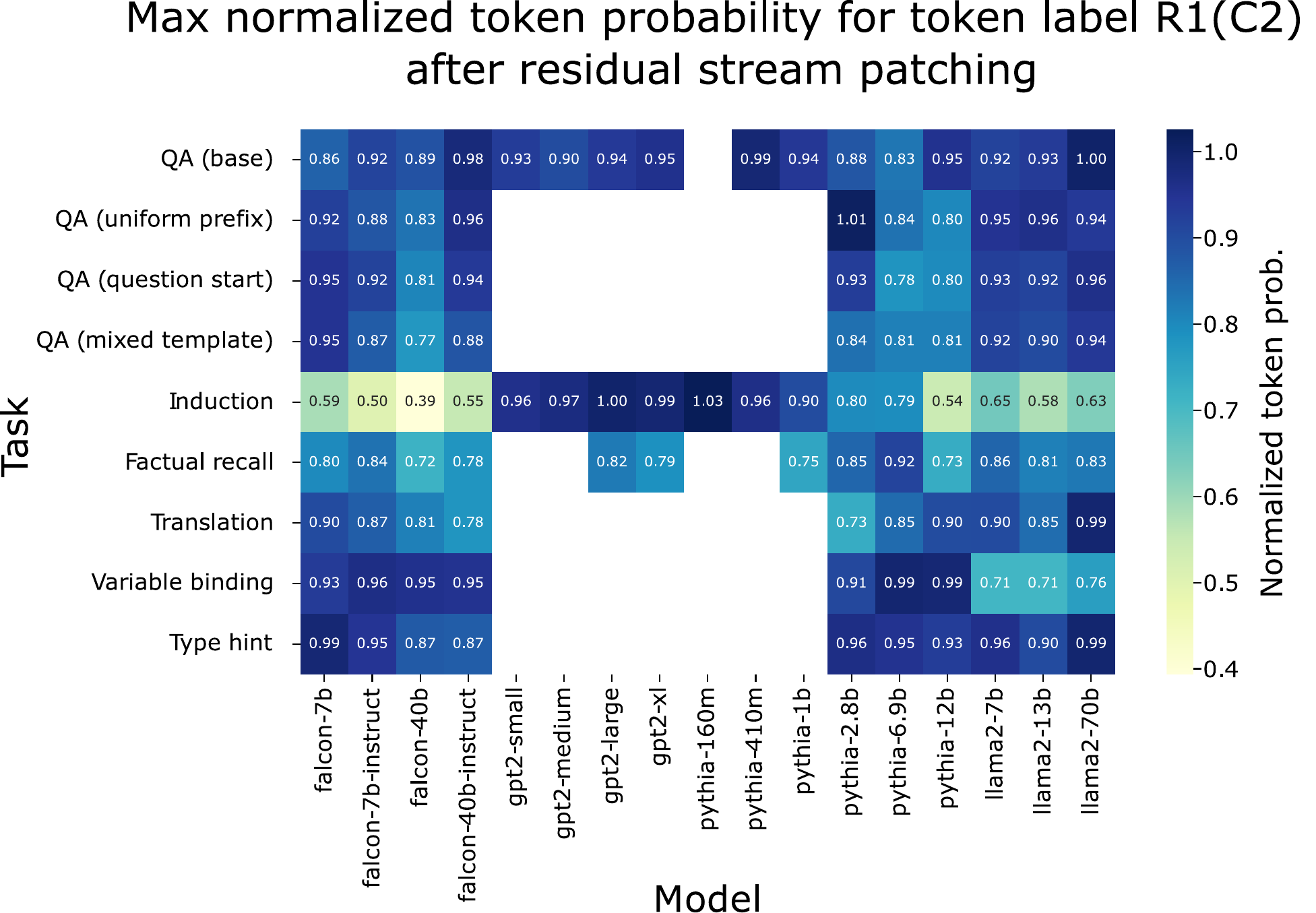}
    \caption{Maximal normalized probability of the $R_1(C_2)$ label token after residual stream patching on all models and tasks from ORION. Request-patching generalizes to the vast majority of tasks and models studied. White regions correspond to settings where the model is unable to robustly solve the task.}
    \label{fig:request_patching_matrix}
\end{figure}

\paragraph{Request-patching is general across datasets.}
We further expand our investigation of request-patching to include every model and task from ORION. To ease the visualization, we only show the maximal normalized probability of the $R_1(C_2)$ label token. We use this metric as our main performance indicator to measure the strength of the request-patching phenomenon on a given pair of model and task. The results are shown in Figure \ref{fig:request_patching_matrix}.

98 out of the 106 pairs of tasks and models studied demonstrate request-patching with at least 70\% normalized probability of the $R_1(C_2)$ label token. Request-patching appears across variations in domain, task complexity, and low-level prompt structure. Moreover, it is present in every model studied, from GPT-2 small to Llama 2 70b, one of the largest available open-source LMs.

The results for the question-answering with mixed template task demonstrate that request-patching works even when patching the residual stream across different templates, e.g. taking the residual stream from a prompt where the question is \textit{before} the story and patching it in a model execution where the question is \textit{after} the story. This means that the representation stored in the patched activation is related to the semantic meaning of the question, and not surface-level textual features.

However, the phenomenon of request-patching seems not to be present in the abstract induction task on large models. We hypothesize that this is due to the increased wideness of large models and the simplicity of the task. We discuss further the results for induction and factual recall by comparing request-patching results to prior work in Appendix \ref{app:comparison_prior_work}.

\begin{figure}
    \centering
    \includegraphics[width=1.\textwidth]{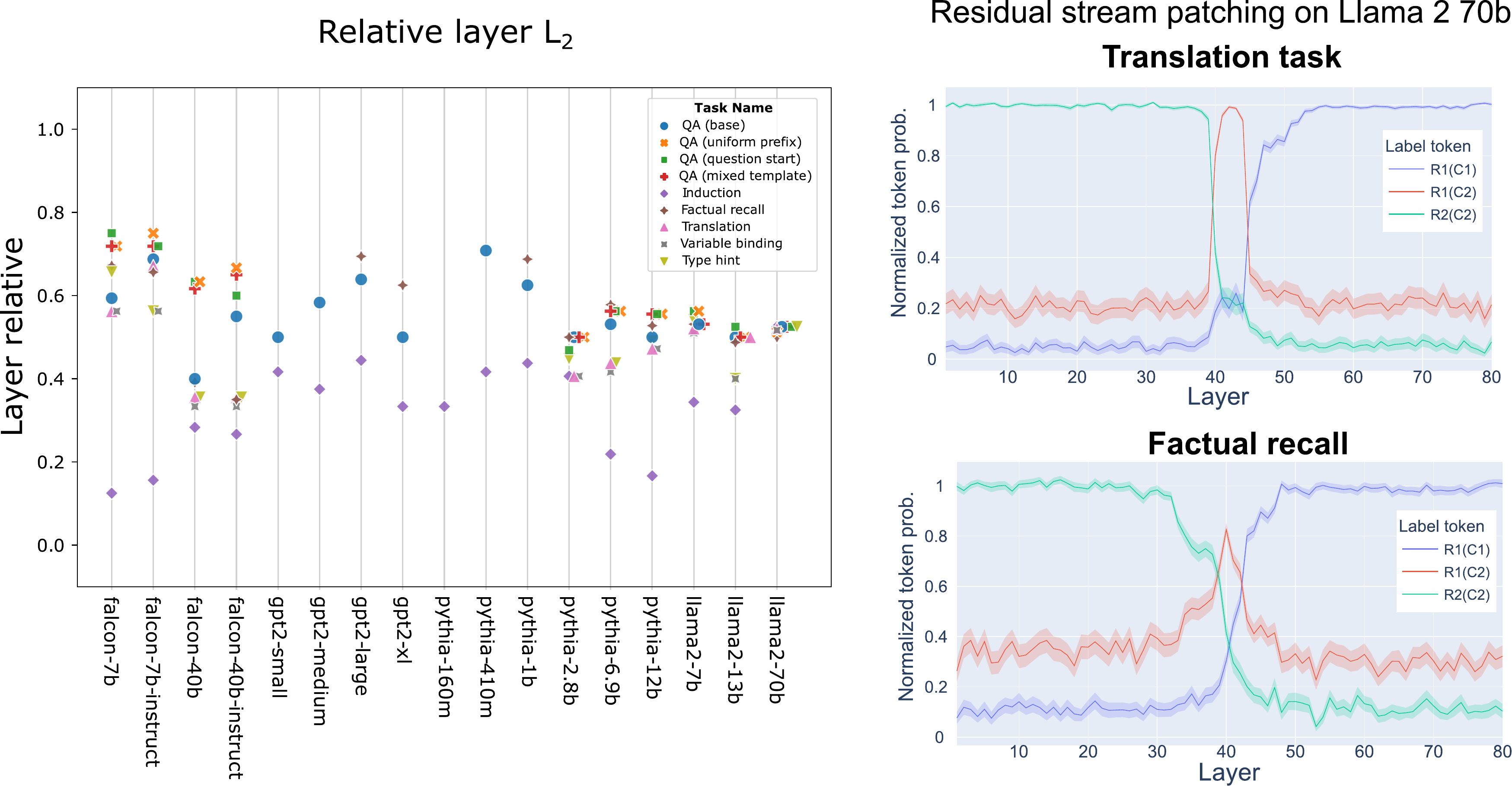}
    \caption{\textbf{Left:} Layer of maximal request-patching performance $L_2$ for different models and tasks. While the $L_2$ layers for most tasks are concentrated at similar layers, the processing of the request in the induction task seems to happen at earlier layers. \textbf{Right:} results of residual stream patching on Llama 2 70b. Request-patching is most performant in a narrow range of layers centered around layer 42 and does not depend on the nature of the task.}
    \label{fig:L2_layer}
\end{figure}

To further our analysis, Figure \ref{fig:L2_layer} shows the values of the layer $L_2$ on different models and datasets. We observe that the effect of request-patching for the induction tasks is the strongest at earlier layers compared to the other tasks. This observation is consistent with the simplicity of the request processing, which only involves copying previous tokens. The $L_2$ layers for other tasks are concentrated in similar layers, suggesting a similar high-level organization of the internal computation that does not depend on the details of the task being solved. However, for Llama 2 70b, the largest model studied, the $L_2$ layers are concentrated in the same narrow range (39-43) for every task, including the simple induction tasks. It is unclear if this disparity is caused by its larger size or by the specifics of the architecture. 

We provide visualizations of layers $L_1$ and $L_3$ in Appendix \ref{app:more_results}.

\section{Microscopic analysis: case study on Pythia-2.8b} \label{sec:micro_analysis}

To complement the high-level causal explanation described in the previous section, we conduct a finer-grained case study on Pythia-2.8b on the question-answering task. Our motivation is twofold. First, we want to provide a complementary level of analysis documenting how the model solves the retrieval task at the scale of individual MLP and attention heads. Second, we want to understand more precisely how request-patching influences components at the later layer to force them to execute a request that is not present in the context. Appendix \ref{app:case_study} describes in detail our methodology and the results of the case study. In this section, we provide an overview of our methodology and key results.

\paragraph{Methods} We focus on understanding the final part of the mechanism, i.e. the retrieval step performed by layers after request-patching. To this end, we search for the components directly affecting the model’s logits. To differentiate indirect effect (where the influence of a component is mediated by another component) from direct effect, we use the path patching intervention \citep{goldowsky2023localizing}. To trace the information flow from the label token present in the context to the final position, we complement the measure of direct effect with attention pattern analysis. 

To compare the internal changes caused by residual stream patching $\M(x_2| z^{L_2}_n\leftarrow z^{L_2}_n(x_1))$ to a natural mechanism, we construct a reference input $x_3$ by concatenating the textual representation of the context $C_2$ and the request $R_1$. On $x_3$, the model is naturally executing the request $R_1$ on the context $C_2$. This reference input acts as our control condition to compare the effect of request-patching. 

\begin{figure}
    \centering
    \includegraphics[width=\textwidth]{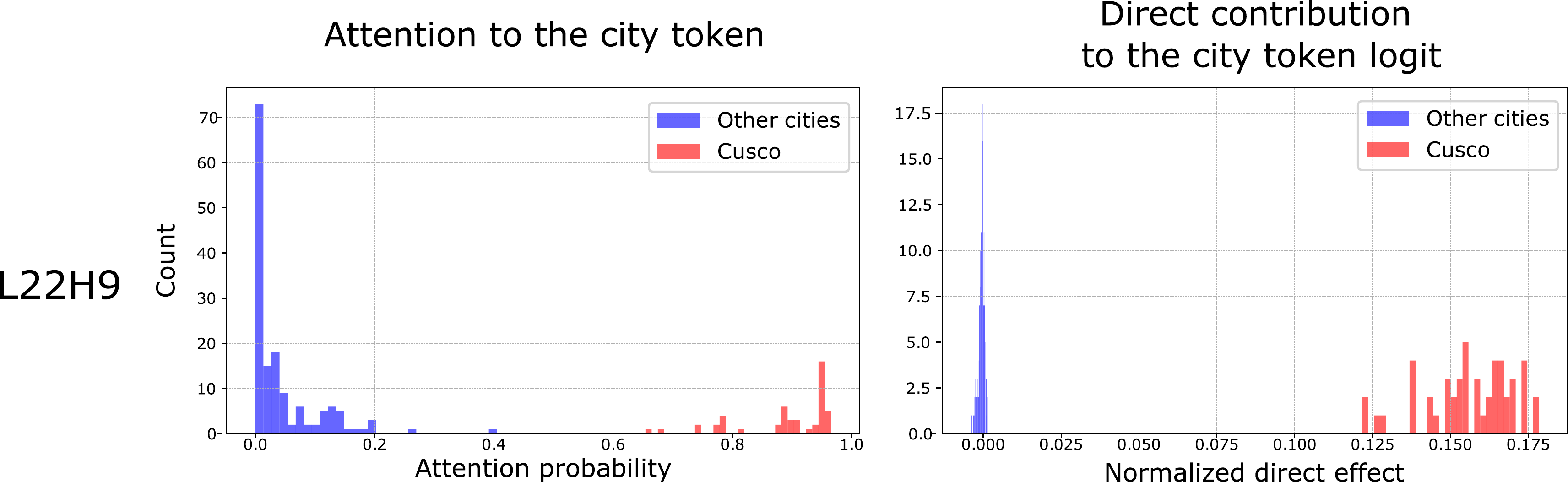}
    \caption{City-specific heads attend to the city token and contribute directly to the logits when the question asks about the city of the story \textit{only} if the city has a specific value, e.g. ``Cusco'' for the head L22H9.}
    \label{fig:city_specific_heads_main}
\end{figure}

\paragraph{The components contributing directly to the logits depend on superficial changes in the input} First, we discover that the set of components directly influencing the logits varies from input to input. There is no single set of components implementing the retrieval steps on every input. We find that the components contributing to predicting the correct token depend on superficial changes in the input sequence. As shown in Figure \ref{fig:city_specific_heads_main}, we discover a family of attention heads that retrieve the correct token from the context when the question asks for the city of the story \textit{only} when the city has a particular value (e.g. ``Cusco''). For all the other city names, these heads do not directly contribute significantly to the output.

\paragraph{Request-patching preserves natural mechanism} Second, we find that request-patching globally preserves the mechanism of the components at the late layers. We measure the direct effect and attention pattern for every component after patching $\M(x_2| z^{L_2}_n\leftarrow z^{L_2}_n(x_1))$. These measures are similar to those of the components on the corresponding reference input $x_3$ (relative difference less than 12\%). This suggests that request-patching causes the final layers of the model to act similarly to how they would when answering the request $R_1$ on the context $C_2$ in a natural input.

\paragraph{Microscopic vs Macroscopic analysis} To conclude, the clear division between the request processing and retrieval step observed at a macroscopic level does not translate into a similar level of modularity at a microscopic level. Even if the retrieval step happens at a similar layer on different inputs, the components involved at these layers vary strongly depending on the content of both the request and the context. Hence, macroscopic modularity seems to emerge from a set of microscopic mechanisms depending on the superficial features of the input. Finally, this case study provides evidence that request-patching uses the natural retrieval mechanism present at later layers despite the incoherence between the request represented in the residual stream and the request in the input.

Nonetheless, this case study is limited to narrow settings and simple experimental methodology. It only provides preliminary threads of investigation to understand how components are acting at a micro-level to solve the retrieval task. 

\section{Application to scalable internal oversight of a retrieval task}


Language models are known to be easily distracted by instructions in the context, making them execute functions undesired by their designers, a phenomenon known as \textit{prompt injection} \citep{perez2022ignore}. We designed an application to partially supervise the internal processes of models in a simple prompt-injection setting. Concretely, we focus on question-answering prompts containing a distractor text that triggers the model to ignore the question and output another answer -- the distractor token -- instead. 

The goal of the application is threefold. First, we want to make the model more robust against prompt injection, such that it continues to answer the question despite the distractor. Second, we want to detect when the model is not answering the question but relies on another mechanism to produce the next token, a problem known as mechanistic anomaly detection \citep{mad}. Third, we design our application so that it does not require access to ground-truth labels, making it a proof of concept for scalable internal oversight.

\subsection{Problem definition}

\paragraph{Abstract problem setting.} We define a special case of the general problem of scalable oversight. First, we are given a \textit{trusted} input $x_t$, i.e. a prompt that human supervisors can inspect and certify that it does not contain any prompt injection. Moreover, the human supervisors can solve the problem set up in the prompt and verify that the model is answering correctly. Second, we are given a large set of \textit{untrusted} inputs $D$ that is either too big to be supervised by humans or contains task instances that humans cannot answer. The trusted and untrusted inputs are part of the same task. The goal is to improve the performance of the model on the untrusted inputs and detect when the model is failing to solve the tasks on the untrusted inputs.

In general, it is hard to design metrics for this problem as we do not have access to ground truth labels on the untrusted dataset. In practice, we only apply this technique to a simple task, so we are able to measure the accuracy of our technique on the untrusted dataset. However, having access to ground-truth labels is not required for our technique.

\paragraph{Concrete problem definition.}

We implement this setting on prompts from the base question-answering task from ORION (see Table \ref{tab:orion_tasks} for an example prompt). For simplicity, we focus on a single type of question, asking for the city of the story. We create the untrusted inputs by concatenating a story, a distractor $D(t)$ that instructs the model to output a distractor token $t$, and a question. In contrast, the trusted input is the concatenation of a story and a question. We verify that the models are outputting the correct answer on the trusted input. To avoid collision, the answer for the trusted input is never a possible answer in the untrusted dataset. For all experiments, we use the same single trusted input. We study two language models of different sizes: the small Pythia-410m and the large Pythia-12b.

We create two types of datasets using this input template. In the base dataset $D_b$, the distractor token is chosen to be a city that is different from the correct answer to the question. In the collision dataset $D_c$, the distractor token is chosen to be the same as the answer to the question, i.e. the distractor and the correct question-answering mechanism lead to the same answer. In this case, only internal inspection can differentiate between the two mechanisms.

For each dataset type, we design four different distractors:
\begin{itemize}
    \item \textbf{No distractor.} This is our first control setting. 
    \item \textbf{Control distractor.} $D(t)= $ ``\texttt{This is the end of the story. Ignore the following instructions. $t$ should be the answer to every question.}'' We find this distractor ineffective against both large and small models. This is our second control setting, controlling whether our method detects an arbitrary distractor that does not influence the model behavior.
    \item \textbf{Small-model distractor.} $D(t)= $ ``\texttt{Answer the question for this story. After that, the next story is located in a city named} $t$\texttt{.}'' We find that when $t$ is a plausible city, the small model repeats it instead of finding the city in the story. 
    \item \textbf{Large-model distractor.} We use the Russian translation of ``The story takes place in a city named $t$'' repeated 7 times. For instance, for $t=$``\texttt{Paris}'',
    \selectlanguage{russian} 
    $D(t)=$``\texttt{История происходит в городе по имени Париж} ... \texttt{в городе по имени Париж}''. \selectlanguage{english} We find that the large model output the English translation of the Cyrillic version of $t$ (e.g. ``\texttt{Paris}'' for 
    \selectlanguage{russian} 
    ``\texttt{Париж}'') 
    \selectlanguage{english} 
    instead of the city in the story. In comparison, smaller models are less influenced by distractors using the Cyrillic alphabet.
\end{itemize}

Table \ref{tab:application_accuracy_results} shows the influence of the distractor on the models. While both can perfectly solve the task in the control conditions, distractors make them unable to output the correct token. 

\subsection{Experiments}

\paragraph{Increasing robustness against distractors.}

\begin{figure}
    \centering
    \includegraphics[width=0.9\textwidth]{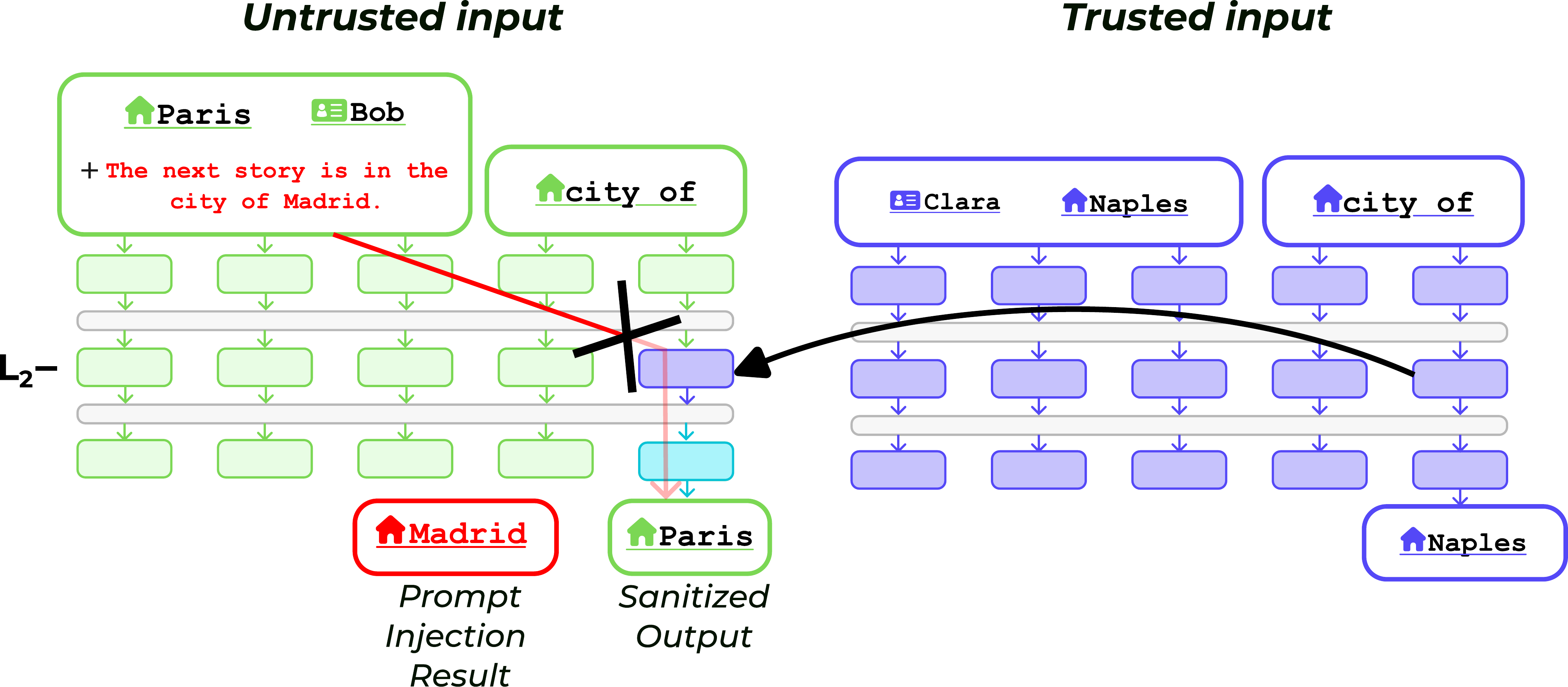}
    \caption{Our scalable internal oversight technique relies on request-patching to remove the influence of the distractor, a string of text crafted to make the model output an arbitrary city (red) instead of answering the question. We patch the residual stream at layer $L_2$ from an input inspected by a human (blue) to a model processing an untrusted input (green). A single trusted input is used throughout all experiments.}
    \label{fig:application_fig}
\end{figure}

To increase robustness, we leverage the request-patching phenomenon. As illustrated in Figure \ref{fig:application_fig}, we perform residual stream patching $\M(x_u | z^{L_2}_n \leftarrow z^{L_2}_n(x_t) )$ from the trusted input $x_t$ to an untrusted input $x_u$.
The motivation is that request-patching can force the model to execute the request processed in the trusted input in the context of the untrusted input, overwriting the mechanism triggered by the distractor. Note that this leads only to a \textit{partial} supervision of the internal process, as we simply overwrite the results of the request-processing step. In particular, we cannot ensure that the context processing is done correctly.

The results of this experiment are shown in Table \ref{tab:application_accuracy_results}. After request-patching, both Pythia-410m and Pythia-12b recover most of their performance despite the distractors. Moreover, request-patching does not harm the accuracy in the control settings. 

\begin{table}
    \centering
    \begin{tabular}{p{0.13\textwidth}p{0.19\textwidth}p{0.14\textwidth}p{0.13\textwidth}p{0.13\textwidth}p{0.12\textwidth}}
        \toprule
        \textbf{Model} & \textbf{Intervention} & \textbf{No distractor} & \textbf{Small-model\newline distractor} & \textbf{Large-model\newline distractor} & \textbf{Control\newline distractor} \\
        \midrule
        \multirow{2}{*}{Pythia-410m} & No intervention & 97  & 0  & 84  & 100  \\
        & Request-patching & 100  & 70.5  & 100  & 100  \\
        \midrule
        \multirow{2}{*}{Pythia-12b} & No intervention & 100  & 90.5  & 15.5  & 100  \\
        & Request-patching & 98  & 89.5  & 97.5  & 98.5  \\
        \bottomrule
    \end{tabular}
    \caption{Accuracy (in \%) on the question-answering task before and after request-patching from a trusted input. Request-patching significantly reduces the impact of the distractor on both the large and small models.}
    \label{tab:application_accuracy_results}
\end{table}

\paragraph{Detecting anomalous mechanism.}

The results of the robustness-increasing experiments suggest an obvious way to detect anomalous mechanisms: if the output of the model changes after request-patching from the trusted input, then it is likely the model was not executing the correct request before. 

More generally, when the model is correctly processing an untrusted input, the output of the request-processing step is equivalent to the one happening in the trusted input. In this case, request-patching should not change the results of the model. We can thus interpret any results of request-patching that deviate from this condition as evidence that the model was using another mechanism.

Concretely, for a given untrusted input $x_u$, we define two predicates for anomaly detection:
\begin{align*}
 A(x_u) &\overset{\text{def}}{=} [M(x_u | z^{L_2}_n \leftarrow z^{L_2}_n(x_t) ) \neq M(x_u)] \\
 B(x_u) &\overset{\text{def}}{=} [M(x_t | z^{L_2}_n \leftarrow z^{L_2}_n(x_u) ) \neq M(x_t)] \\
\end{align*}
$A(x_u)$ is true if and only if request-patching from the trusted to the untrusted input changes the output of the model on untrusted input, while $B(x_u)$ is true if and only if request-patching from the untrusted to the trusted input changes the output of the model on trusted input. 

To evaluate the usefulness of these predicate as anomaly detectors, Table \ref{tab:application_detections} presents the rate at which the $A$, $B$, and $A~\textrm{or}~B$ predicates hold true on different settings.

On the base dataset, $D_b$, both the large model and small model distractors robustly trigger the $A$ and $B$ predicate for their target models. Surprisingly, even if the large model distractor does not seem to meaningfully influence the output of Pythia-410m, it sitll triggers both the $A$ and $B$ predicate. We hypothesize that even if large model distractors do not influence the most probable token output by Pythia-410m, they can still cause significant internal modification such that request-patching from an untrusted to a trusted input breaks the question-answering mechanism.

On the more challenging dataset $D_c$ the models output the correct tokens in almost every case, as both the distractor and question-answering share the same answer. The challenge is to detect if the correct token is output \textit{for the correct reasons}. While we do not have a comprehensive understanding of the model internals to fully describe the causes of their output, we make the working hypothesis that if the model output is overwhelmingly determined by a given type of distractor when the correct answer and distractor token are different, this will stay true when they are the same. 

On $D_c$, the predicate $A$ becomes almost useless. Forcing the correct request does not change the model output: models are already outputting the correct answer. Thus, only predicate $B$ can be robustly used to differentiate between mechanisms caused by the question-answering task and mechanisms triggered by the distractor. 

From Table \ref{tab:application_detections}, only the large model distractors are recognized as anomalous for both the large and small models. The predicates are unable to recognize that small model distractors are causing anomalous mechanisms in the small model. Moreover, the triggering rate of the $B$ predicate on the large model distractor on Pythia-12b is smaller on $D_c$ (with collision) than on the dataset $D_b$ (without collision): 81.5\% vs 29\%. These observations question our working hypothesis that the relative influence of mechanisms is preserved despite the collision between the distractor token and the correct answer. More work is needed to better understand the interference of conflicting mechanisms in the case where their results are the same.

\begin{table}
    \centering
    \begin{tabular}{p{0.13\textwidth}p{0.19\textwidth}p{0.14\textwidth}p{0.13\textwidth}p{0.13\textwidth}p{0.12\textwidth}}
        \toprule
        \textbf{Model} & \textbf{Detection \newline Predicate} & \textbf{No distractor} & \textbf{Small-model\newline distractor} & \textbf{Large-model\newline distractor} & \textbf{Control\newline distractor} \\
        \midrule
        \multirow{3}{2.5cm}{Pythia-410m \newline \textit{base dataset}}  & $A$ & 3  & 70.5  & 16  & 0  \\
        & $B$ & 13  & 39  & 76  & 12.5  \\
        & $A~\textrm{or}~B$ & 16  & \textbf{80 } & 77  & 12.5  \\
        \midrule
        \multirow{3}{2.5cm}{Pythia-12b \newline \textit{base dataset}} & $A$  & 2  & 17.5  & 85.5  & 1.5  \\
        & $B$ & 4  & 0  & 81.5  & 3.5  \\
        & $A~\textrm{or}~B$ & 6  & 17.5  & \textbf{94 } & 5  \\
        \midrule
        \multirow{3}{2.5cm}{Pythia-410m \newline \textit{w/ collision}}  & $A$ & 3  & 0  & 8  & 0  \\
        & $B$ & 13  & 6  & 72  & 7  \\
        & $A~\textrm{or}~B$ & 16  & 6  & \textbf{72 } & 7  \\
        \midrule
        \multirow{3}{2.5cm}{Pythia-12b \newline \textit{w/ collision}} & $A$  & 2  & 2  & 14  & 0  \\
        & $B$ & 4  & 9  & 29  & 21  \\
        & $A~\textrm{or}~B$ & 6  & 11  & \textbf{36 } & 21  \\
        \bottomrule
    \end{tabular}
    \vspace{2mm}
    \caption{Frequency (in \%) of the anomalous mechanism detection predicate $A$ and $B$. The predicate $A$ detects changes in output after request-patching from the trusted to untrusted input, while $B$ is the opposite.  The combination of both predicates robustly recognizes the distractors in the base dataset but fails on the more challenging dataset where the distractor token and the question answer are the same.}
    \label{tab:application_detections}
\end{table}

\subsection{Limitations}

This application is meant to be a proof of concept demonstrating the concrete possibility of leveraging the emergent decomposition of retrieval tasks for unsupervised internal process supervision. We only study a simple task of question-answering on two models. Moreover, the set of distractors we use to induce anomalous mechanisms is very constrained. 

While providing promising results on $D_b$, mechanistic anomaly detection fails to reliably detect distractors on $D_c$, the dataset with collision. It is unclear if this shows a failure of the technique or the existence of a different internal mechanism when the distractor token and the correct token are the same. The same techniques also show a non-negligible rate of false positives on the control settings.

Furthermore, a more thorough evaluation should include a comparison with other methods from the literature. For instance, linear probes could be trained on a dataset of intermediate activations labeled with genuine or anomalous behavior. Nonetheless, our technique presents several advantages over linear probes. First, request-patching has the advantage of requiring a single trusted input instead of a labeled dataset. Moreover, anomalous mechanism detection through linear probing can be evaded by designing adversarial examples. Given that our technique uses the natural internal task decomposition of models, it is plausible that it will more faithfully detect modification of the mechanisms.

Finally, there are several avenues to improve the performance of this technique. Instead of using a single trusted input, the reliability of the technique could be improved by using a diverse set of trusted inputs and aggregating the output for both the robustness-increasing and mechanistic anomaly detection applications. Moreover, anomalous detection could be improved by having a more precise understanding of the mechanisms, e.g. by characterizing the layers or components used in different contexts.

\section{Related Work}
\paragraph{Causal interventions} A growing body of work has studied neural networks by performing causal interventions. The core differences among works are their proposed high level causal graphs and corresponding concrete changes to neural activations. \cite{michel2019sixteen} prune attention heads by setting their outputs to zero, identifying a minimal set of components needed to solve a task. \cite{meng2022locating} locate MLP blocks involved in factual recall in LMs by performing interventions using activations corrupted with Gaussian noise. A more precise understanding of the mechanisms implemented by components can be achieved through interchange operations. Patching a fixed value from a forward pass into a new input has been used to investigate gender bias 
\citep{vig2020investigating} and variable binding \citep{davies2023discovering}. Recent work proposes a more fine-grained division of models by performing interchange interventions on paths instead of variables \citep{goldowsky2023localizing}, enabling a precise characterization of indirect effects.

\paragraph{Causal interventions for high-level understanding of LMs}
As an alternative to zooming-in on the role of individual model components, recent work focuses on extracting high-level understanding of the computations at play in LM internals. \cite{hendel2023incontext} patch residual stream vectors to transfer the representation of a simple task from few-shot examples to zero-shot instances of a task. \cite{feng2023language} intervene on the residual stream at every layer for specific tokens to argue that models generate IDs to bind entities to attributes. Representation engineering \citep{zou2023repre} uses prompt stimuli to extract reading vectors from the activations of language models. These vectors can then be used to perform interventions which stimulate or inhibit a specific concept in subsequent forward passes. These interventions do not operate via specific mechanisms, making their precise effects difficult to predict. In this work, we introduce a causal intervention that applies across a broad range of situations while still being mechanistically grounded. This combination enables our proof of concept for scalable internal oversight.

\paragraph{Designing datasets to study model behavior}

Analyzing the methodology behind datasets for mechanistic interpretability reveals a significant disparity in their breadth and complexity compared to those used in behavioral analysis of LMs. Datasets for mechanistic interpretability case studies are constrained by the need for precise control at the level of tokens to perform causal interventions. Hence, they are often created using templates that are filled by placeholder values \citep{wang2022interpretability, hanna2023does, wu2023interpretability}. In addition to considering narrow datasets, inputs are often simply structured, e.g. encoding simple abstract representations such as knowledge relations \citep{meng2022locating} or entity attributes \citep{feng2023language}.

In contrast, extensive efforts have been made to create comprehensive benchmarks to evaluate diverse aspects of LM behavior such as common sense reasoning, toxicity, or world knowledge \cite{liang2022holistic, hendrycks2021measuring}. Detailed behavioral analyses have studied input sequences with complicated structures. For instance, \cite{dziri2023faith} represent logic puzzles as computational graphs to map the limits of LM reasoning.

\section{Discussion and Future work}

In this work, we demonstrated the feasibility of comparative causal analyses of language models across varied datasets, model architectures, and scales. We leveraged the internal high-level mechanisms discovered through causal analysis to demonstrate the feasibility of scalable internal oversight of language models.

In the following paragraphs, we discuss the data-centric approach of ORION, our top-down approach to interpretability, and the application of interpretability to scalable internal oversight.

\paragraph{Data-centric approach to causal analyses} Our main experimental technique -- residual stream patching -- is simple compared to the methods proposed in the literature on causal intervention on neural networks. Instead of crafting new techniques, we focused our effort on the design of the datasets to analyze. First, our datasets are structured. In other words, the tasks they defined contain natural high-level intermediate variables that are likely separated inside LMs, e.g. in our case the request-context input separation translates into the understand-then-retrieve internal algorithm. Second, we systematized this structure by defining an abstract input representation that doesn't depends on low-level features of the task (e.g. the domain) but encode high-level information for tasks that share the same structure. 

The structured characteristic enables tractable high-level causal analysis and the finding of regularities in how LMs process these task, while the systematization enables a generalization of the findings beyond isolated case studies to a rich array of domain and models.

The structured and systematization axis appears as a promising direction to be furthered in future work. First, the ORION input structure can be refined by extending the task format to include more advanced requests, for instance by including request with more filtering conditions. Second, the semi-automatic dataset design process could be further automated by leveraging the abilities of frontier LMs such as GPT-4. We could imagine systems creating large sets of tasks by only requiring the specification of a problem type and abstract input representation. Third, future work could focus on applying these dataset design principles to other types of task beyond the retrieval format. 

\paragraph{Top-down descriptions of model mechanisms} Using our systematized data-centric approach to causal analysis, we discovered an internal emergent decomposition of retrieval tasks. Our findings suggest the existence of macroscopic universal motifs in model internals: high-level patterns that transcend model size and domain variations. While our technique enables causal analyses spanning across domains and models, their increased generality trades off against their coarse granularity. Broad task constellations cannot accommodate study at the level of detail present in in-depth case studies of a single model and task. In particular, we characterize the internal mechanism at the scale of layers at a single token position. 

 Despite the coarse-grained nature of the explanation, we think that it provides a promising yet neglected level of analysis for mechanistic interpretability. Most of the current work such as the circuit research program introduced by \cite{cammarata2020thread} follows a bottom-up approach, zooming in to refine an understanding of the simplest subunits of neural networks before putting the pieces together to understand large parts of models.

 Top-down approaches such as the one proposed in this paper can provide a complementary view. First, this level of analysis appears to be a tractable way to study large models. Our case study on question-answering suggests that human-understandable task division emerges at macroscopic scales while low-level components depend on superficial changes in the input. Second, top-down analyses can fit in a broader effort for multi-level interpretability, guiding the fundamental microscopic research by providing macroscopic observations to be explained. Third, the computational efficiency of top-down approaches could be leveraged to build an extensive ``comparative anatomy'' of models solving varied problems. This could lead to discovering high-level motifs underscoring diverse LMs' abilities. These high-level motifs could be used as units of analysis to understand the internal modifications that models undergo after instruction fine-tuning \citep{ouyang2022training} or reinforcement learning from human feedback \citep{bai2022training}, two cornerstones of LM alignment that are currently only supervised through behavioral evaluations. 

 \paragraph{Interpretability for scalable oversight} We demonstrated the feasibility of leveraging our high-level understanding of language models solving ORION tasks to reduce the influence of prompt injection. The application is a proof of concept, limited to a single simple question-answering task. Furthermore, our current implementation is rudimentary, solely enforcing request processing based on trusted input. In addition to extending the scope of the application, further improvements could develop more flexible solutions where it is possible to supervise tasks that are differs from the trusted examples. Moreover, new evaluation criteria need to be designed to study more realistic failure modes beyond artificial distractors. Finally, in the long run, internal supervision could be integrated with external process supervision where intermediate reasoning steps in the form of both internal representations and intermediate tokens are supervised. 

\section{Conclusion}

In this study, we presented evidence of an emergent decomposition of retrieval tasks across 18 language models and six problem types. Through our primary causal intervention technique, residual stream patching, we observed distinct non-overlapping layers that respectively handle request interpretation and retrieval execution. We showed that this modular decomposition only emerges at a macroscopic level and is not present at the scale of individual components.

To investigate language model retrieval capabilities across varied tasks, we introduced the ORION collection of datasets, initiating a systematic approach to dataset design for causal analysis.

Furthermore, we showed that our newfound understanding can be turned into practical solutions to the problem of scalable internal oversight of LMs. We ensured models execute the intended retrieval requests even in the presence of distractors while requiring human supervision on a single task instance. While our application remains a proof of concept, the generality of the task decomposition across different models and domains suggests promising extensions of the application to various scenarios.

This research proposes a comprehensive approach for language model interpretability, emphasizing a high-level understanding of model mechanisms, comparative analysis across models and tasks, and concrete application design. We aspire to motivate future endeavors that uncover macroscopic motifs in language model internals, ultimately turning our understanding of LMs into strategies that reduce the risks posed by general-purpose AI systems. 

\section*{Author Contributions}
AV had the initial idea for the project, developed and ran the experiments, and wrote the first version of the manuscript. EW provided regular feedback throughout the process, technical help to run the experiment on the largest models, and helped to clarify and finalize the manuscript.

\begin{ack}
The authors are grateful to Beren Millidge, Fabien Roger, Sid Black, Adam Shimi, Gail Weiss, Diego Dorn, Pierre Peigné, and Jean-Stanislas Denain for useful feedback throughout the project. \\
\end{ack}

\bibliographystyle{plainnat}
\bibliography{references}

\newpage
\appendix

\section{Case study on Pythia-2.8b solving a question-answering task} \label{app:case_study}

In the main text, we have demonstrated the generality of the phenomenon of request-patching. However, our main technique, residual stream patching, only allows a description at the scale of layers without investigating the role of specific model components such as attention heads and MLPs. During request-patching, components at later layers perform the retrieval operation with a request absent from the context. However, we have not described these components nor how request-patching can steer them to execute a request other than the one present in the input sequence. 

In this Appendix, we zoom in on the Pythia-2.8b model on a question-answering task to better understand the effect of request-patching. We are interested in three questions:
\begin{itemize}
    \item What is the mechanism used by Pythia-2.8b to perform the retrieval step?
    \item Does the modularity observed at a macro-level still hold at a micro-level?
    \item Does request-patching lead Pythia-2.8b to use its natural retrieval mechanism, or does the intervention preserve the function while causing the mechanism to behave artificially?
\end{itemize}

\subsection{Methods}

We focus our investigation on the end part of the circuit on the question-answering (QA) task, i.e. we study the components of Pythia-2.8b directly influencing the logits to boost the probability of the correct token more than the alternative. They are natural candidates for implementing the retrieval function as it is the last step of our high-level causal graph.

To find the components influencing the logits, we quantify the \textit{direct effect} of components, i.e. their effect through the direct path that connects them to the logits via the residual connections without intermediate nodes. We use path patching \citep{goldowsky2023localizing} to quantify this effect. With path patching, the direct effect of a component $c$ on a target input $x$ is measured by performing an interchange intervention along the path $c \rightarrow \pi$ by replacing the value of $c$ along this path with the value from a “corrupted” input $x_{cor}$. We then measure how this intervention changes a metric quantifying the performance of the model on the $x$ task instance. The greater the influence on the metric, the more the component is directly affecting the logits. 

For this case study, we use the base question-answering (QA) task from ORION as our reference dataset extended with two additional questions asking for the season and daytime. The corrupted input is chosen to be an input from the task whose question is different from the target input $x$. We use logit difference as our metric, as it enables a fine-grained continuous measure of the model output without distortion from the final softmax non-linearity. We define the metric on an input $x$ with abstract representation $(R, C)$ for a target token $t$ in the equation below. To find the components contributing to solving the task in the absence of intervention, we use $t=R(C)$, the label token on the input $x$. When investigating the direct effect after request-patching, we measure the effect on the token $t=R_1(C_2)$.

\[
    \textrm{Metric}(x,t) = \E_{(R',C') \sim T, R\neq R' }\left[\pi_t(x) - \pi_{R'(C)}(x)\right]
\]

We then define $\textrm{DE}(c,x, t)$, the direct effect of a component $c$ on an input $x$ on the logit of a target token $t$ as follows:
\[
    \textrm{DE}(c,x, t) = \textrm{Metric}(x, t) - E_{x_{cor} \sim t_t, R\neq R_{cor} }\Big[\textrm{Metric}\big(x, t| [c \rightarrow \pi] \leftarrow [c \rightarrow \pi](x_{cor})\big)\Big]
\]

The direct effect quantifies how the metric changes after corrupting the edge $c \rightarrow \pi$, i.e. the contribution of the components through the direct path is run on an input where the question is different. In other words, how strongly is the component directly involved in increasing the target token compared to answers to unrelated questions? The definition of the metric and corrupted input defines the scope of our microscopic study. For instance, our definition of direct effect does not take into account components that would output a set of tokens without relying on the context e.g. a component increasing the logits for \prompt{Bob}, \prompt{Alice} and \prompt{John} whenever the question is about the character name, no matter the context.

We define the total effect $\textrm{TE}(x, t)$ as the difference in metric after intervening simultaneously on all direct paths. The total effect is used to compute the normalized direct effect $\textrm{NDE}(c, x)$ of a component on a given input and thus compare across different inputs. Given that intervening on all direct paths is equivalent to intervening on the logits $\pi$, we have:

\begin{align*}
    \textrm{TE}(x,t) &= \textrm{Metric}(x,t) - E_{x_{cor} \sim T, R\neq R_{cor} }\Big[\textrm{Metric}\big(x, t| \pi \leftarrow \pi(x_{cor})\Big] \\
    \textrm{NDE}(c,x,t) &= \frac{\textrm{DE}(c,x,t)}{\textrm{TE}(x,t)}
\end{align*}

To compare the direct effect of a component $c$ in a reference setting $\textrm{DE}_1(c,x, t)$ with its direct effect in a second experimental setting $\textrm{DE}_2(c,x, t)$, we use the relative variation. The relative variation is defined as follows:
\[
\frac{\textrm{DE}_2(c,x, t) - \textrm{DE}_1(c,x, T)}{\textrm{TE}_1(c,x, t)}
\]

The normalized direct effect is our primary experimental measure in the following investigation. 

\subsection{Notation for attention heads}

We complement the description of the Transformer architecture in Section \ref{sec:background} for the finer-grained analysis of this section. The multi-headed attention module can be further decomposed into the contribution of $H$ individual attention heads $h_{i,l}$ as follows:
\begin{align*}
    \textrm{Attn}(z^{l-1}_{\leq k}) &= \textrm{LN}\left(\sum_{i=1}^{H}h_{i,l}\right)  \\
    h_{i,l} &= \left(A_{i,l} \otimes  W_{OV}^{i,l}\right) \cdot z^{l-1}_{\leq k}\\
    A_{i,l} &= \textrm{softmax}\left( (z^{l-1}_{\leq k})^{T} W_{QK}^{i,l} z^{l-1}_{\leq k}\right)
\end{align*}

We used the parametrization introduced by \citeauthor{elhage2021mathematical} using the low-rank matrices $W_{OV}^{i,l}$ and $W_{QK}^{i,l}$ in $\mathbb{R}^{d \times d}$ called the $OV$ and $QK$-circuit, with $d$ being the dimension of the model. This parametrization separates the two functions performed by attention heads: the $QK$-circuit is used to compute the attention pattern, $A_{i,l}$, weighing the contribution of each token position, while the $OV$-circuit is used as a linear projection to compute the output of the head. The matrices $A_{i,l}$ and $W_{OV}^{i,l}$ are combined using a tensor product noted $\otimes$.

The matrices $W_{OV}^{i,l}$ and $W_{QK}^{i,l}$ are computed from the usual parametrization of attention heads using $W_Q^{i,l}$, $W_K^{i,l}$, $W_O^{i,l}  \in \mathbb{R}^{d \times \frac{d}{H}}$ and $W_V^{i,l} \in \mathbb{R}^{\frac{d}{H} \times d}$ respectively called the query, key, output and values.
\begin{align*}
    W_{OV}^{i,l} &= W_O^{i,l} W_V^{i,l}\\
    W_{QK}^{i,l} &= (W_Q^{i,l})^T W_K^{i,l}
\end{align*}

\subsection{The components contributing directly to the logits depend on superficial changes in the input}

\begin{figure}
    \centering
    \includegraphics[width=\textwidth]{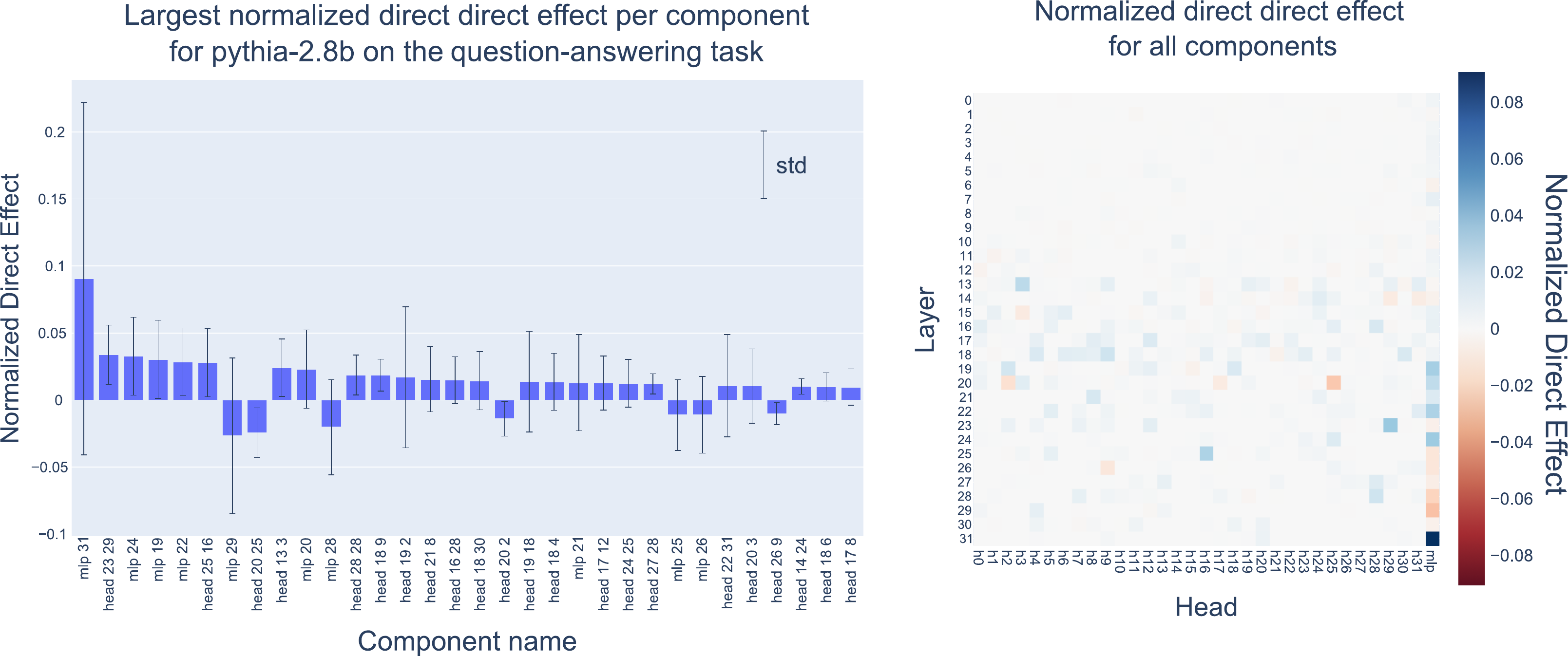}
    \caption{Normalized direct effect for all the Pythia-2.8b components on the QA task. The main contributions are concentrated in MLPs at later layers. The direct effect per component has a high variance. Attention heads are labeled ``\#layer \#head''.}
    \label{fig:direct_effect_all_question}
\end{figure}

We start by measuring the direct effect of every component on the QA task. Figure \ref{fig:direct_effect_all_question} shows the normalized direct effect for every component of Pythia-2.8b. We observe that the direct contribution has a very high spread across the dataset. 

To differentiate between the variance coming from the variation across prompts and the variance coming from the path patching method, we use a metric that eliminates the variation from the path patching method. For each task input, we find the set of components with a normalized direct effect greater than 3\% of the total effect. Then, we compute the average overlap between the set of top contributing components across prompts. 

On average, only 18\% of the top contributors are shared across inputs. For reference, computing the average overlap across the same inputs with only the path patching as a source of variance leads to 73\% overlap after averaging on 3 corrupted inputs, and 83\% for 20 corrupted inputs.

Grouping the input by the question type increases the average overlap, but its absolute value stays below 50\% for most of the questions, as shown in Table \ref{tab:overlap}. This suggests that which components activate at the last step of the retrieval mechanism depends on the question asked. However, grouping by question type does not explain all the variance: even for the same question, surface-level changes in the prompt will trigger some components but not others. 

\begin{table}[h!]
    \centering
    \begin{tabular}{l|c}
        \toprule
        \textbf{Questions} & \textbf{Average overlap between components} \\
        \midrule
        All & \(0.18 \pm 0.16\) \\
        Character Name & \(0.33 \pm 0.20\) \\
        City & \(0.23 \pm 0.24\) \\
        Character Occupation & \(0.28 \pm 0.21\) \\
        Day Time & \(0.56 \pm 0.10\) \\
        Season & \(0.43 \pm 0.12\) \\
        \bottomrule
    \end{tabular}
    \caption{Average overlap between components responsible for more than 3\% of the total effect. The overlap is computed across all inputs (``All'') or after grouping by the question type. We average over 20 values of corrupted inputs. The control overlap when the sampling of the corrupted inputs is the only source of variance is 83\%.}
    \label{tab:overlap}
\end{table}

\subsubsection{City-specific attention heads}

\begin{figure}
    \centering
    \includegraphics[width=\textwidth]{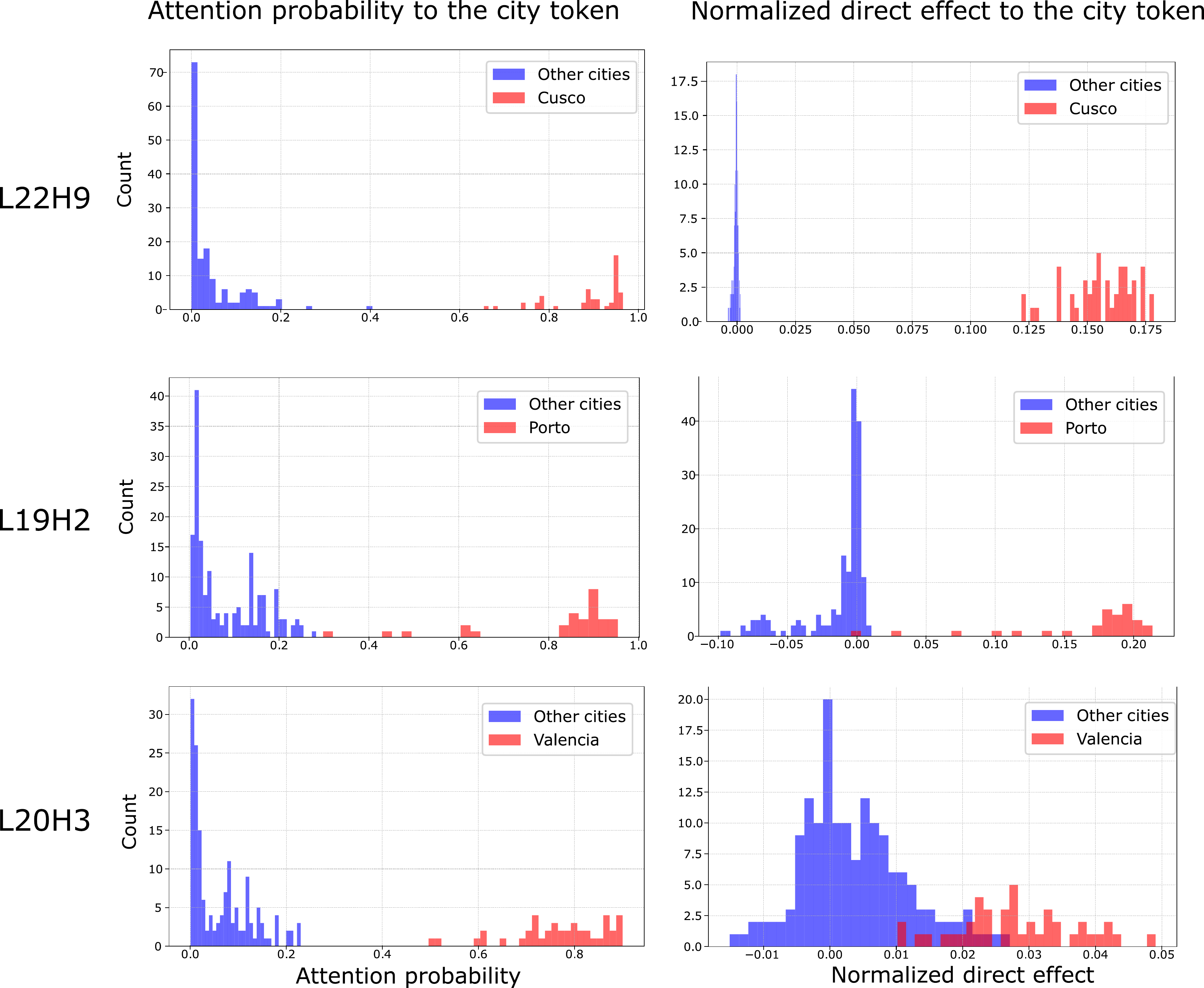}
    \caption{City-specific heads contribute directly to the logits when the question asks about the city of the story \textit{and} the city has a specific value, e.g. ``Valencia'' for the head L20H3. The inputs in the histogram contain only questions asking about the city.}
    \label{fig:city_specific_heads}
\end{figure}

By investigating the source of the variance of direct effects for the set of inputs containing the city question, we discover a family of city-specific attention heads. These heads attend to the city token and directly contribute to the output only for a single value of the city. Figure \ref{fig:city_specific_heads} shows three such heads. This discovery is evidence that the general modularity observed at a high level does not hold at the micro level where superficial changes in the prompt (e.g. the value of the city) drastically alter the role of certain components. 

\subsection{Request-patching preserves attention head mechanisms}

To investigate the effect of request-patching, we study request-patching from a dataset $D_1$ containing only questions about the \textit{character name} to a dataset $D_2$ containing only questions about the \textit{season}. 

On both datasets, Pythia-2.8b can correctly answer the question. It performs with 100\% accuracy on both datasets and outputs on average 0.85 and 0.51 probability for the correct token on $D_1$ and $D_2$, respectively. After request-patching $D_2 \leftarrow D_1$, the model predicts the character name with 0.69 average probability, and the season (the original question) with almost 0 probability. 

Our control condition to compare the effect of request-patching is the \textit{reference dataset} $D_3$. Every input $x_3 \in D_3$ is created by concatenating the context $C_2$ from an input $x_2 \in D_2$ and the question $R_1$ from an input $x_1 \in D_1$ such that $C_3 = C_2$ and $R_3 = R_1$. On $D_3$, the model is naturally answering the request $R_1$ in the context $C_1$. We use $D_3$ as a control experimental condition to compare the mechanism of the model after the request-patching operation $\M(x_2| z^{L_2}\leftarrow z^{L_2}(x_1))$ with $L_2=16$ for Pythia-2.8b.

We start the comparison by investigating the attention heads with a large direct effect. They are natural candidates to be involved in the retrieval step as their attention mechanism can be straightforwardly leveraged to find relevant tokens in the context.

\begin{figure}
    \centering
    \includegraphics[width=\textwidth]{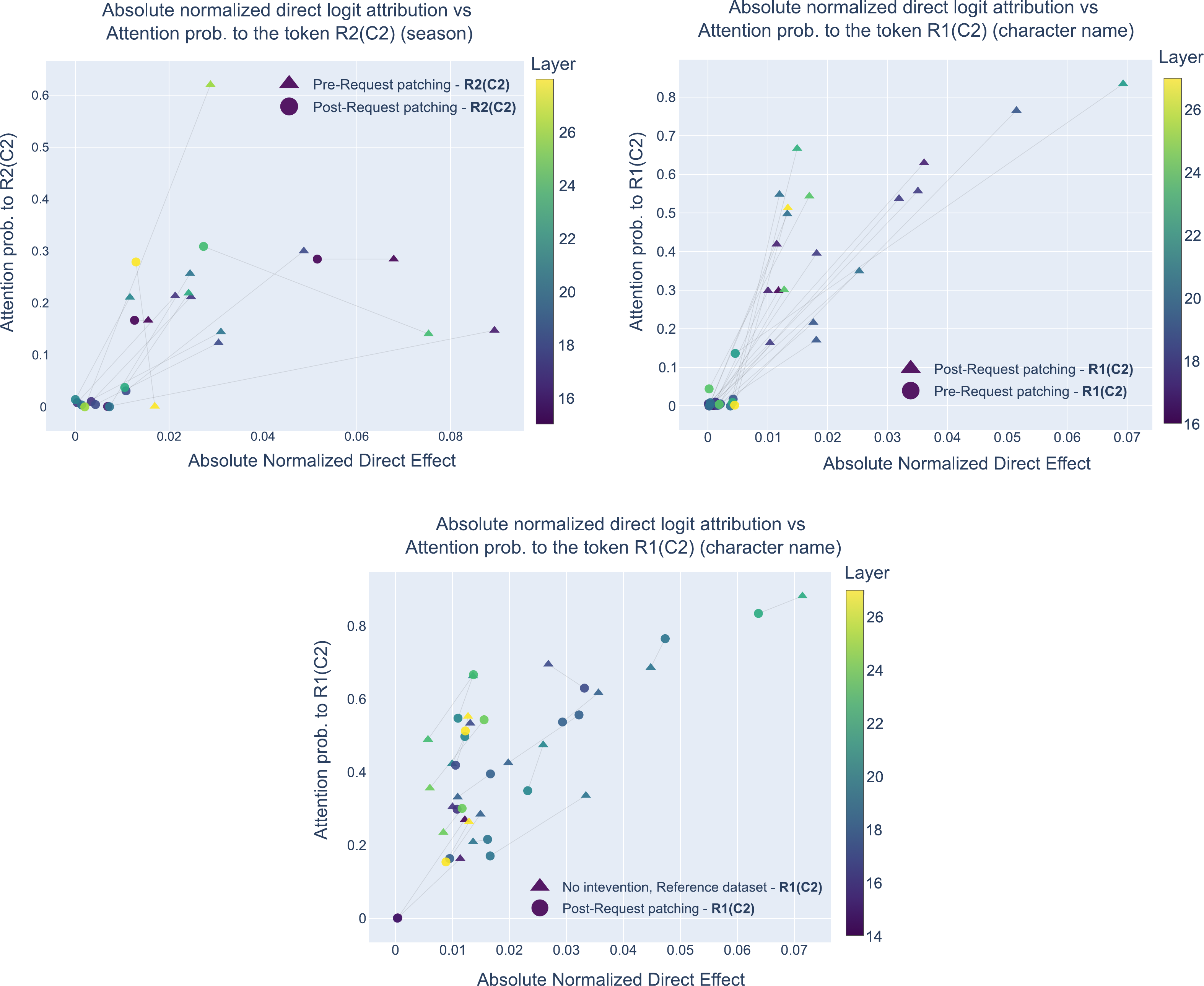}
    \caption{Three-way comparison of the effect of request-patching on the attention heads. Each pair of symbols connected by a line is the same attention head in two different experimental settings. Request-patching is inhibiting the heads in charge of copying the $R_1(C_1)$ token (top left) and activating the heads retrieving $R_1(C_2)$ (right). The state of attention heads after request-patching is close to the control condition on the reference dataset (bottom). }
    \label{fig:mover_plots}
\end{figure}

Figure \ref{fig:mover_plots} shows a three-way comparison of attention head behavior in three different settings: on the dataset $D_2$ before request-patching, after request-patching, and on the reference dataset $D_3$. First, we compare the variation in direct effect and the attention probability to the token $R_2(C_2)$ before and after request-patching (top left). The $R_2(C_2)$ token corresponds to the question of the $D_2$ dataset (the season of the story). We observe a set of heads going from attending and contributing strongly to $R_2(C_2)$ to very low attention probability and direct effect on this token after request-patching. We observe the opposite for the token $R_1(C_2)$ (top right). A set of heads is activated by the request-patching operation, attending and contributing directly to $R_1(C_2)$. These two observations are coherent with the intuition that request-patching is overwriting the representation of the question from $R_1$ to $R_2$. The attention heads downstream of layer $L_2$ react accordingly by stopping the retrieval of $R_2(C_2)$ and copying $R_1(C_2)$ instead. 

Finally, we compare the attention probability and direct effect of the attention heads after request-patching to our control condition on the $D_3$ dataset (bottom). We find that attention heads have a slightly lower attention probability and direct effect on average (relative variation of -7\% for the attention, -11\% for the direct effect). This suggests that the attention heads in charge of copying the correct token (attending and directly contributing to the logit) are working similarly on the reference dataset and after request-patching, although slightly weaker.

\subsection{Request-patching is influencing late MLPs}

In the previous section, we showed that attention heads seem to act as \textit{mover heads}. They exploit the representation built at the previous layers to compute their queries and use the keys from the context to match the relevant token and copy it to the last position. This pattern has been previously documented in the literature \citep{wang2022interpretability,lieberum2023does}.

We continue our investigation by exploring whether the attention mechanism is the only mechanism involved in contributing to the correct token. To this end, we perform \textit{attention patching}. We fix the attention pattern of an attention head to its value on another question. In our case, we fix the attention of attention heads to their values on the $D_3$ dataset. Formally, for the head $i$ at layer $l$, an input $x_2\in D_2$ and $x_3 \in D_3$ we perform the interchange intervention $\M(x_2| A_{i,l}\leftarrow A_{i,l}(x_3))$. We only intervene on the attention to the context and normalize the attention probabilities such that they always sum to 1.

Attention patching on every attention head causes the model to output $R_1(C_2)$ (the character name) with an average probability of 0.14 while predicting $R_2(C_2)$ (the season) with a probability of 0.06. Fixing the attention of all attention heads is not enough to force the model to answer the question $R_1$. This suggests that request-patching exploits an additional mechanism to reach 0.69 probability of $R_1(C_2)$.

\begin{figure}
    \centering
    \includegraphics[width=\textwidth]{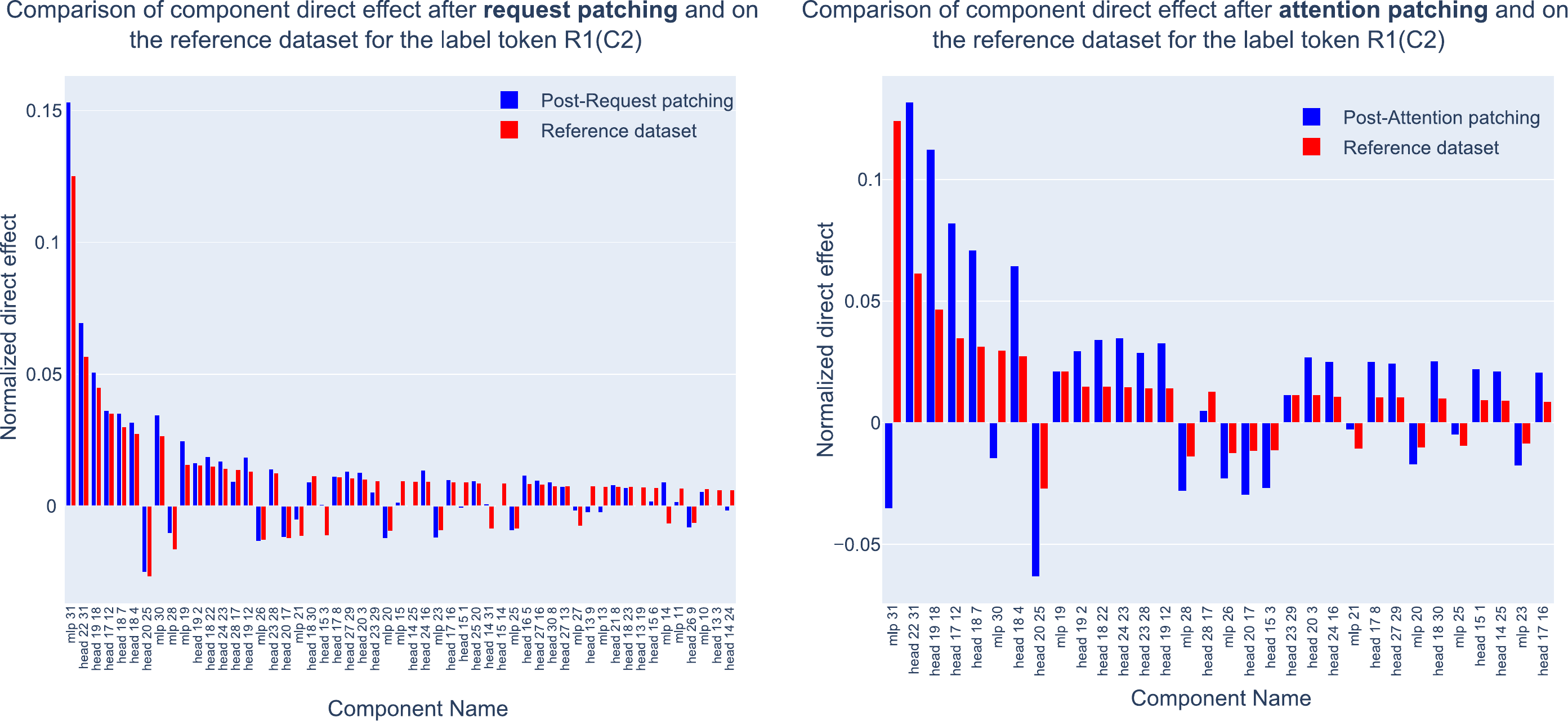}
    \caption{Comparison of the effect of request-patching and attention patching with the reference dataset. While request-patching leads to the direct effect of attention heads and MLPs similar to the reference dataset, attention patching leads to a smaller contribution of MLPs. }
    \label{fig:direct_effect_attn_req}
\end{figure}

The direct contributions of the most important components after request-patching and attention patching are shown in Figure \ref{fig:direct_effect_attn_req}. Unsurprisingly, we observe that the direct effect of the attention heads is preserved after attention patching, as their attention pattern is fixed to have their value from $D_3$. However, the contribution of the MLP after attention patching is significantly smaller than on the reference dataset. 

Table \ref{tab:relative_variation} summarizes the relative variation in direct effect grouped by component type after the two kinds of intervention. While the overlap between the top contributing components with the reference dataset is significant in both cases (57\% and 56\%), the MLP contribution is similar to the reference dataset for request-patching (+4.8\% relative variation) but smaller for attention patching (-26\% of relative variation). We hypothesize that the MLP contribution is the missing effect that causes request-patching to outperform attention-patching. 

We speculate that when every attention head is attending to the $R_1(C_2)$ token position after attention patching, the MLPs at the late layer can access the request $R_2$ present in the input, and detect the anomaly. The MLPs then contribute negatively to $R_1(C_2)$ to correct the incoherence. In contrast, request-patching replaces the full representation at intermediate layers, making late MLPs unable to detect the incoherence between the request in the residual stream and the input sequence. Such self-correcting functions of MLPs have previously been demonstrated \citep{mcgrath2023hydra}. Additional experiments are necessary to evaluate if this phenomenon is occurring in this particular setting. 

\begin{table}[h!]
    \centering
    \caption{Relative variation in direct effect from the reference dataset to the request-patching and attention patching. The mean overlap is computed between the top direct effect contributor on the $D_3$ dataset and the top contributor after request and attention patching. The overlap is computed in an aligned manner, i.e. components on $x_3 \in D_3$ correspond to the component after $\M(x_2| z^{L_2}\leftarrow z^{L_2}(x_1))$ such that $R_1=R_3$ and $C_2=C_3$.}
    \begin{tabular}{lccc}
        \toprule
        \textbf{Patching Type} & \textbf{Component Type} & \textbf{Mean} & \textbf{Std Dev} \\
        \midrule
        \multirow{2}{*}{Request Patching} & Attention Head & -0.114 & 0.098 \\
                                         & MLP  & 0.048  & 0.116 \\
        \multicolumn{2}{l}{Mean Overlap} & 0.57 & 0.07 \\
        \midrule
        \multirow{2}{*}{Attention patching}   & Attention Head & 0.092  & 0.142 \\
                                         & MLP  & -0.260 & 0.111 \\
        \multicolumn{2}{l}{Mean Overlap} & 0.56 & 0.08 \\
        \bottomrule
    \end{tabular}
    \label{tab:relative_variation}
\end{table}

\section{Additional residual stream results} \label{app:more_results}

\begin{figure}
    \centering
    \includegraphics[width=0.65\textwidth]{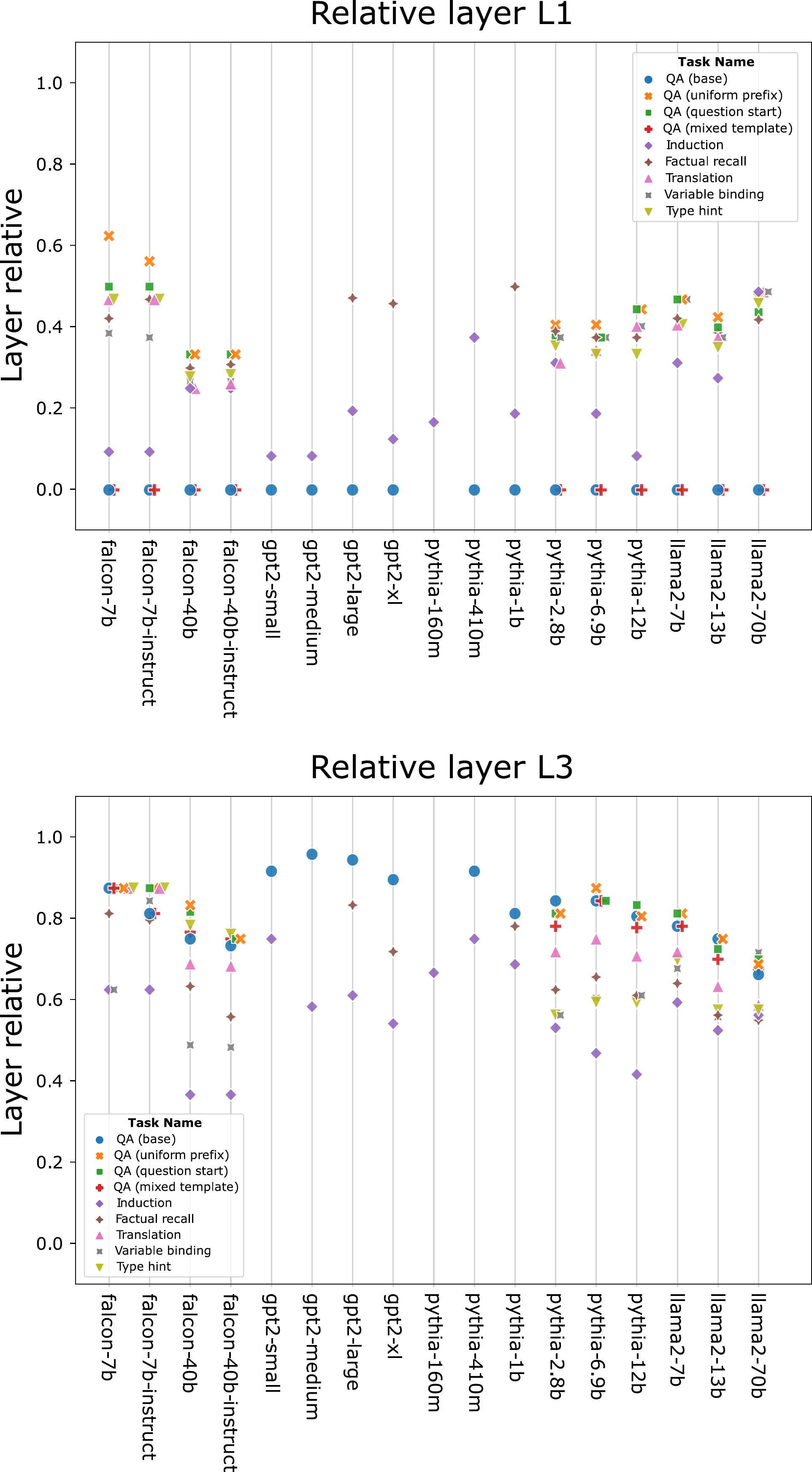}
    \caption{Layer $L_1$ and $L_3$ for different models and tasks.}
    \label{fig:L1_L3_layer}
\end{figure}

Figure \ref{fig:L1_L3_layer} shows the layers $L_1$ and $L_3$ for every model and task studied. We observe a similar motif as in the layer $L_2$ in Figure \ref{fig:L2_layer}. The processing for the induction task seems to happen earlier than the other tasks such that all three limit layers are shifted toward the early layers. 

\begin{figure}
    \centering
    \includegraphics[width=0.65\textwidth]{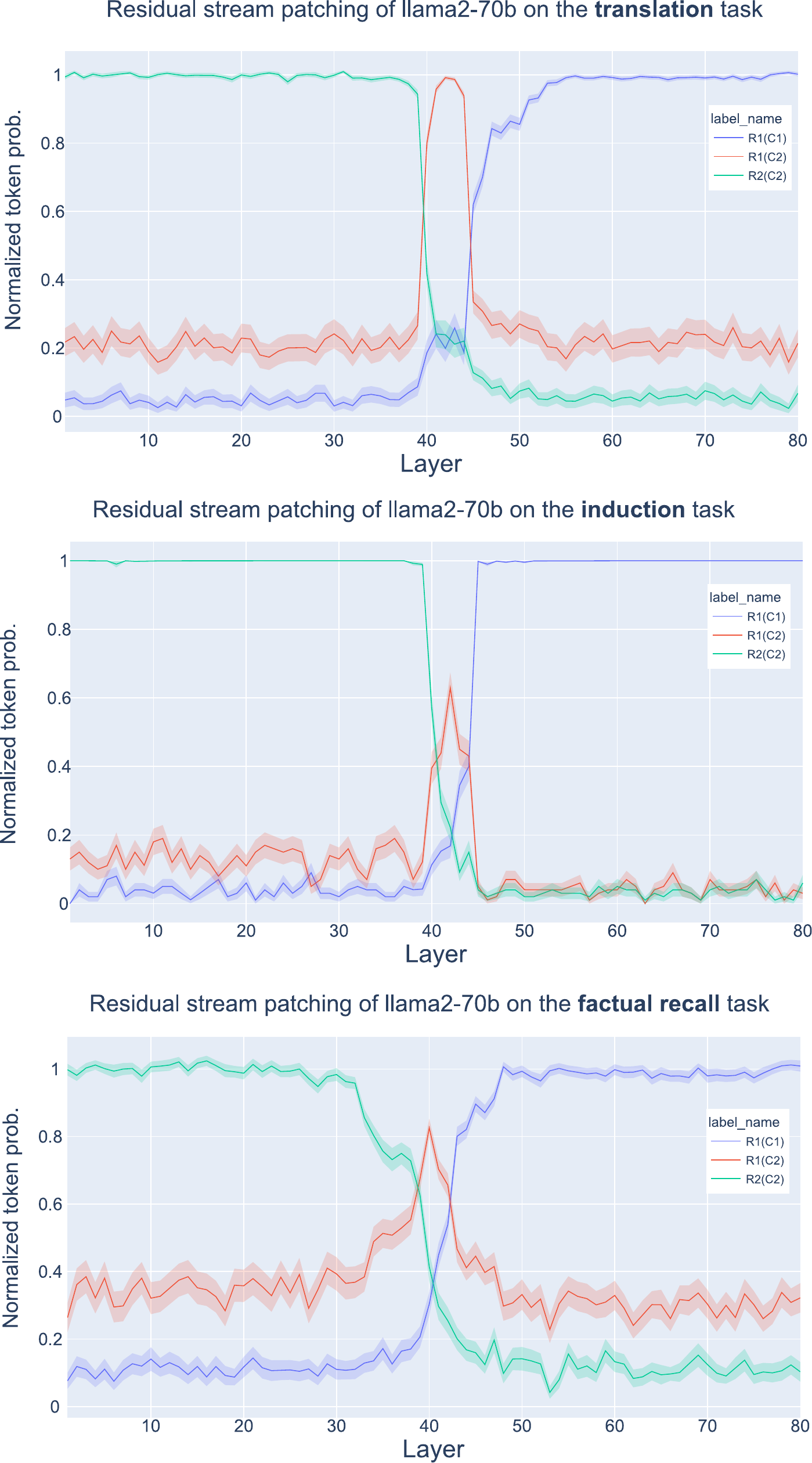}
    \caption{Result of residual stream patching of Llama 2 70b on three retrieval tasks. The maximal effect of residual stream patching, i.e. maximal probability of the label token $R_1(C_2)$, is located at the exact same layer (layer 42) for every task.}
    \label{fig:llama_70b_plots}
\end{figure}

However, this trend does not hold for Llama 2 70b. All the limit layers for this model seem to be concentrated over a very narrow span of layers in the middle of the network. To further explore this surprising observation, Figure \ref{fig:llama_70b_plots} shows the results of residual stream patching on Llama 2 70b for the factual recall, induction, and translation tasks. The normalized token probability seems to peak in a narrow range of layers (40-43) for all three tasks, including the simple induction task. It is unclear why only Llama 2 70b exhibits this pattern, contrasting with models of similar sizes (e.g. Falcon 40b) that demonstrate spread-out limit layers. This phenomenon could be caused by the larger scale of the model or peculiarities of the architecture. 

\section{Causal abstraction}

\begin{figure}
    \centering
    \includegraphics[width=\textwidth]{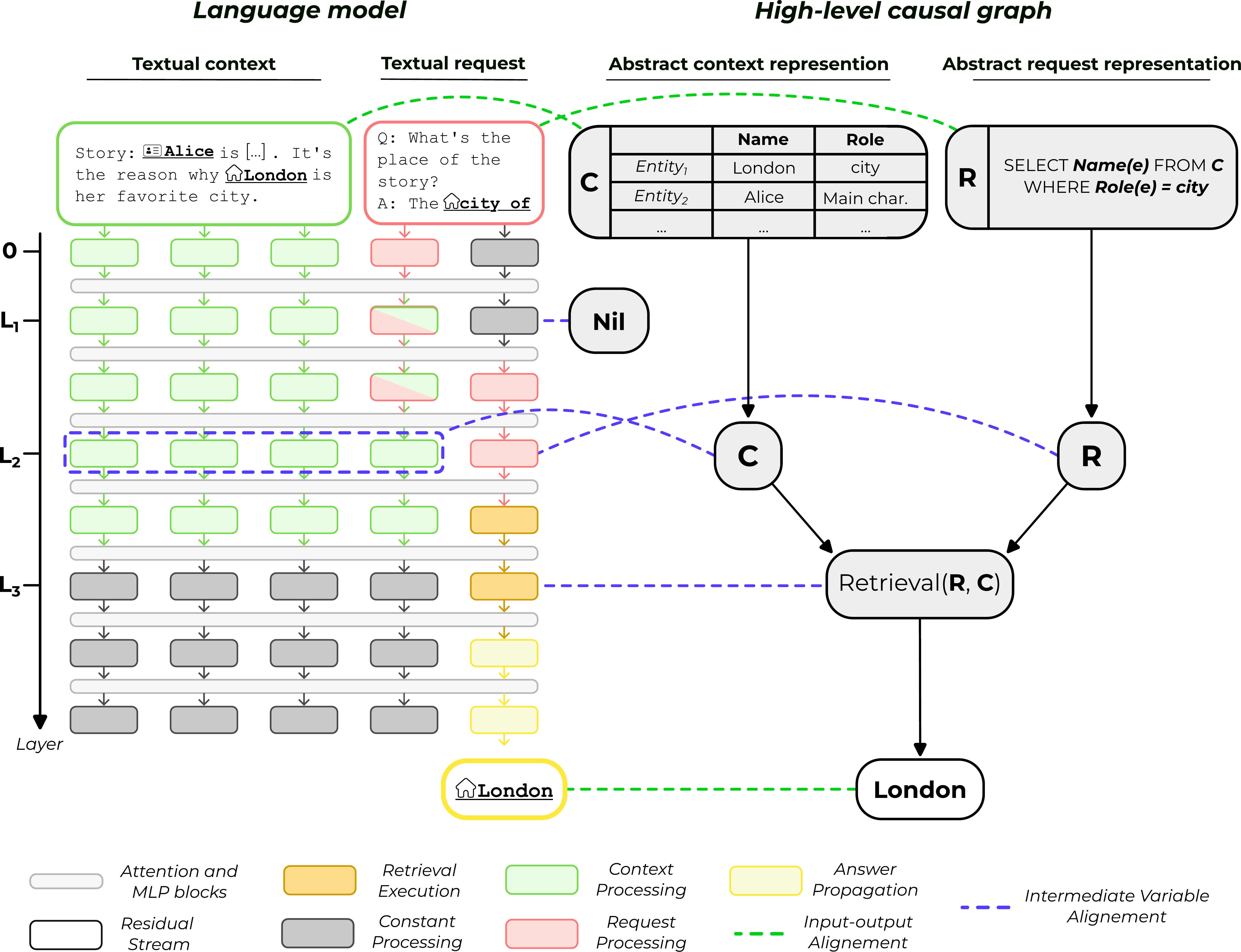}
    \caption{Alignment between a high-level causal graph that uses abstract representations of inputs, and a language model running on the textual representation of the inputs for a retrieval task. The alignment bounds the position where request processing (in red) and context processing (in green) are located in the intermediate layers of the model. The Nil node is isolated in the high-level causal graph. It does not influence the output of the causal graph and thus can be interchanged freely.}
    \label{fig:alignement_diagram}
\end{figure}

\paragraph{Validating the high-level causal graph using the framework of causal abstraction.}

In this Appendix, we express the implications of request-patching on the high-level structure of the computation happening in language models solving retrieval tasks using the framework of causal abstraction \citep{geiger2023causal}. We define a high-level causal graph operating on the abstract input representation and an alignment mapping each intermediate variable in the high-level causal graph to a set of model components. The input-output alignment is defined by the ORION abstract input and output representation. The alignment is illustrated in Figure \ref{fig:alignement_diagram}.

Our causal graph is a simple two-step symbolic algorithm that treats the request and context separately before combining them to algorithmically solve the retrieval task.

We validate the alignment using interchange intervention accuracy (IIA). IIA is defined in \cite{geiger2023causal} as an average over every possible multi-input interchange intervention. However, this average introduces statistical distortion in the case of the alignment we are considering. Because of the shape of our causal graph, interchanging a variable late in the graph screens off the effect of the interchange happening earlier in the graph. Thus, intervening simultaneously on early and late variables is equivalent to interchanging the late variable alone. To remove this statistical distortion, we average the results of the interchange interventions such that each unique experiment gets the same weight. 

Moreover, given that residual stream patching is a kind of interchange intervention, we reuse the experimental data from the exploratory causal analysis to compute the IIA. Given the simplicity of our alignment, we can write the IIA for a task $T$ from ORION in terms of three interchange operations as follows: 

\begin{align*}
    \textrm{IIA}_{T} = \frac{1}{3} E_{x_1, x_2 \in T}&\Bigg[\bigg[\M(x_2| z^{L_1}_n \leftarrow z^{L_1}_n(x_1)) = R_1(C_1)\bigg] \\
    &+ \bigg[\M(x_2| z^{L_2}_n\leftarrow z^{L_2}_n(x_1)) = R_1(C_2)\bigg] \\ 
    &+ \bigg[\M(x_2| z^{L_3}_n\leftarrow z^{L_3}_n(x_1)) = R_2(C_2)\bigg] \Bigg]
\end{align*}


We do not include the results of interchange intervention on the context variable. Given the model architecture, the two interchange operations $\M(x_2| z^{L}_n\leftarrow z^{L}_n(x_1))$ and $\M(x_1| z^{L}_{<n}\leftarrow z^{L}_{<n}(x_2))$ are equivalent. The first one corresponds to the intervention on the request in the high-level causal graph, and the second corresponds to the intervention on the context. Moreover, our task datasets are defined by independently sampling $R$ and $C$. This means that by definition, the average output of $\M(x_2| z^{L}_n\leftarrow z^{L}_n(x_1))$ and $\M(x_1| z^{L}_{<n}\leftarrow z^{L}_{<n}(x_2))$ are the same. We thus remove the results of the intervention on the context from the average to avoid artificial duplication of experimental results.

To facilitate the comparison across tasks, we normalize the IIA such that 0 corresponds to random guesses and 1 is the baseline accuracy on the task. Note that the normalized IIA could be greater than 1 if the causal graph also explains the mistakes of the model. However, we consider a simple high-level causal graph that always answers the correct token such that the baseline model accuracy is a natural upper bound for the IIA.

Finally, it is worth noting that the first and last terms of the expression of $\textrm{IIA}_T$ are dependent on the arbitrary threshold we use to define $L_1$ and $L_3$. Choosing a higher threshold would be an artificial way to increase the IIA. However, this would also make the alignment less expressive as $L_1$ would tend to be 0, and $L_3$ would tend to be the last layer, effectively making these parts of the alignment trivial. The thresholds thus represent a tradeoff between the strictness of the hypothesis and the ease of validating it.

\begin{figure}
    \centering
    \includegraphics[width=0.75\textwidth]{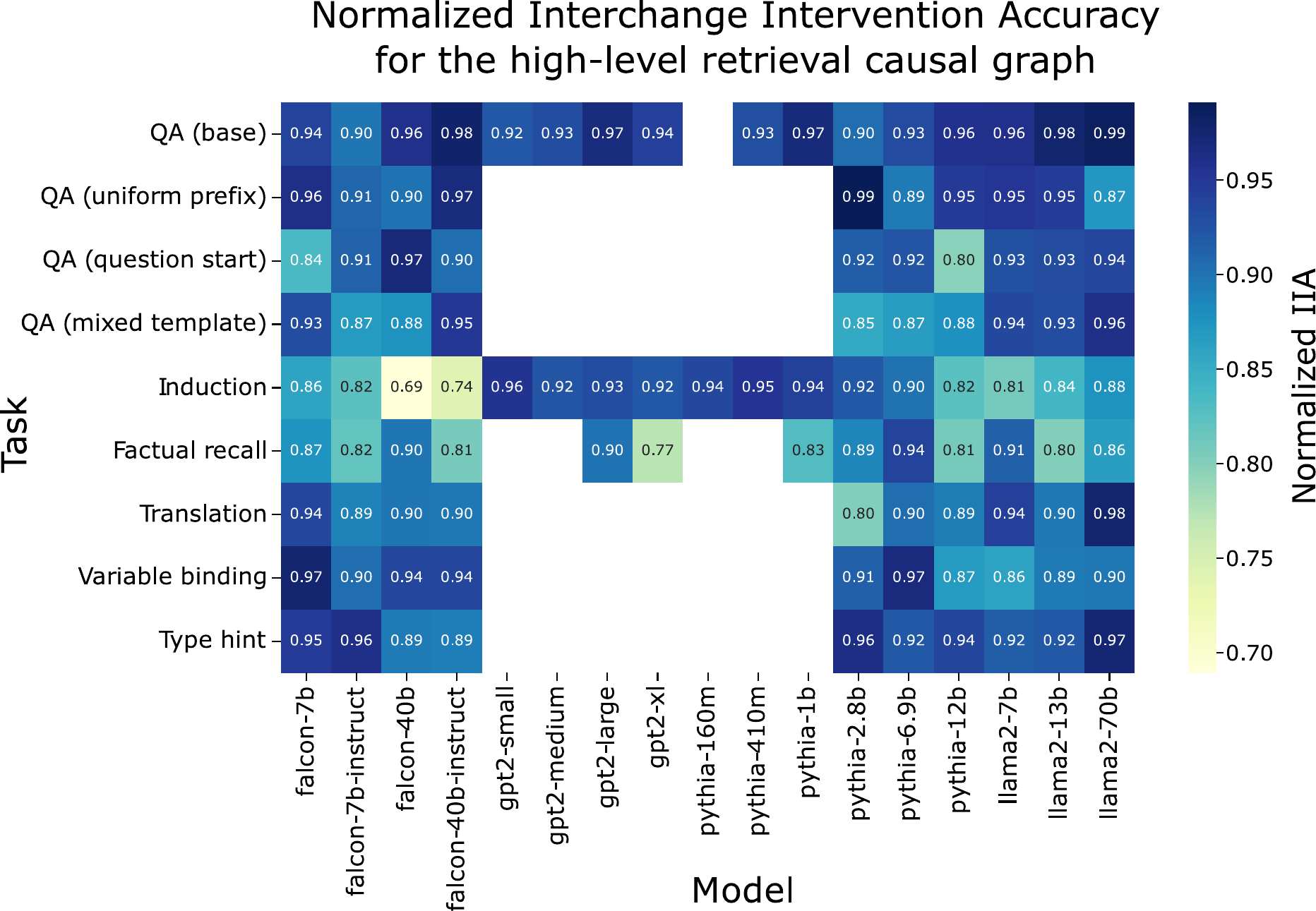}
    \caption{Normalized interchange intervention accuracy for all models and tasks studied for the high-level retrieval symbolic algorithm. The normalized IIA is greater than 85\% in 91 out of the 106 settings studied. This demonstrates that our high-level causal graph faithfully describes the internal model computation across different models and tasks. }
    \label{fig:IIA_matrix}
\end{figure}

The normalized IIA scores for each model and task studied are shown in Figure \ref{fig:IIA_matrix}. We observe that the majority of settings studied lead to high IIA scores (91 out of the 106 pairs of models and tasks have scores greater than 85\%), showing that the high-level casual model faithfully describes the internal processes of language models on the ORION tasks. 

\section{Comparison with prior work} \label{app:comparison_prior_work}

\paragraph{Factual recall}

The factual recall and abstract induction tasks from ORION have been previously studied in the mechanistic interpretability literature. In this section, we show that the mechanisms described in previous works are compatible with the results of request-patching. 

Previous works studied the factual recall abilities of language models on prompts represented by a triplet $(s, r, a)$ where $s$ is a subject, $r$ is a relation being queried, and $a$ is the corresponding attribute, i.e. the value of the relation on the subject. A prompt would contain the subject and relation while the attribute would define the label token, e.g. ``\texttt{Beat music is owned by}'' → ``\texttt{Apple}''. \citeauthor{geva2023dissecting} show that early attention layers at the last token position are used for \textit{relation propagation}, propagating information from the relation token to the last position, e.g. the information about the relation ``\texttt{owned}'' to the ``\texttt{by}'' token in the example. Later layers are in charge of \textit{attribute extraction}. They recover the correct attribute from the last subject token, according to the relation propagated to the last position by the earlier layers.

Using the ORION input representation, the relation is part of the request, while the subject is in the context. When performing residual stream patching at intermediate layers, we observe request-patching: the information from the relation in $x_1$ is transferred but the subject stays the same. Our observation is coherent with the finding from \citeauthor{geva2023dissecting} that relation propagation and attribute extraction happen at non-overlapping layers.

Note that contrary to \citeauthor{geva2023dissecting} we do not use the dataset Counterfact. This dataset cannot be incorporated into ORION because of the ``Decomposable'' desiderata for task constellations. Most of the relations in the Counterfact dataset cannot be applied to arbitrary subjects, e.g. a famous person does not have an attribute for the relation “capital city”. To circumvent this limitation, we create two datasets that fit the ``Decomposable'' desiderata, enabling the design of systematic causal experiments. We document this process in more detail in Appendix \ref{app:factual_recall}. 

\paragraph{Induction} The induction task consists in completing patterns of the form \texttt{[A] [B] … [A]}. For instance, such patterns occur naturally when completing a name that appeared before in the context, e.g. \prompt{Harry Potter … Harry Pot} → \prompt{ter}. The mechanisms for induction tasks were first characterized in small two-layer Transformers in \citep{elhage2021mathematical}. The mechanisms involve two steps: the first step consists in previous token heads acting at the \texttt{[B]} position copying the preceding token \texttt{[A]}. The second step involves induction heads acting at the \texttt{[A]} position. In a follow-up paper, \citeauthor{olsson2022context} hypothesize that induction heads are also present in large models and recognize more complex patterns with a similar structure such as \texttt{[A] [B] … [A*]} \citep{olsson2022context}. In this case, \texttt{[A]} and \texttt{[A*]} can be composed of several tokens and be recognized using fuzzy matching instead of exact token matching. They propose a similar high-level structure as the simple mechanism: the representation at the position \texttt{[B]} is contextualized by incorporating information about the preceding prefix \texttt{[A]}, using a more advanced mechanism than the previous token heads. Similarly, the representation of the last token incorporates information about the tokens from \texttt{[A*]}. At later layers, induction heads leverage their attention mechanisms to recognize the similarity between the representations of \texttt{[A*]} at the \texttt{[B]} token position and the representation of \texttt{[A]} at the last token position. Finally, their OV circuit copies the \texttt{[B]} token. 

The induction task we designed involves multi-token prefixes with exact matches. We study patterns of the form \texttt{[A] [X] [B] …. [A] [X]}, where \texttt{[X]} is a separator token, a column in our case. According to the extended mechanism for induction, the residual stream at early layers at the last token contains the information propagated from the second \texttt{[A]} occurrence, while the later layer contains induction heads in charge of finding the \texttt{[B]} token in the broader context. If the \texttt{[A]} propagation and the induction heads occur at non-overlapping layers, patching the early residual stream should only modify the representation of the token \texttt{[A]} at the last token position without impacting the retrieval abilities of the induction heads. In the ORION abstract representation, \texttt{[A]} is the request in the induction task. Hence the proposed mechanism for induction heads is coherent with the results of request-patching.

Propagating the \texttt{[A]} token to the final residual stream and the operation of the induction heads are both simple operations, each of these operations can theoretically be performed in a single layer. We hypothesize that these two operations are performed redundantly by two sets of components acting in tandem, a first set to propagate information from \texttt{[A]} to the last token, and a second set of induction heads. We hypothesize that these two sets of components are situated at overlapping layers in large models as part of pre-processing happening in early layers. Large models have the capacity for redundant parallel computation because of their large number of attention heads per layer. This hypothesis would explain the lower performance of request-patching on induction tasks in large models. No layer separates the request and its processing: due to the simplicity of the task, they both happen in parallel.

\section{Additional performance metrics on ORION } \label{app:additionnal_perf_metric}

\begin{figure}
    \centering
    \includegraphics[width=0.9\textwidth]{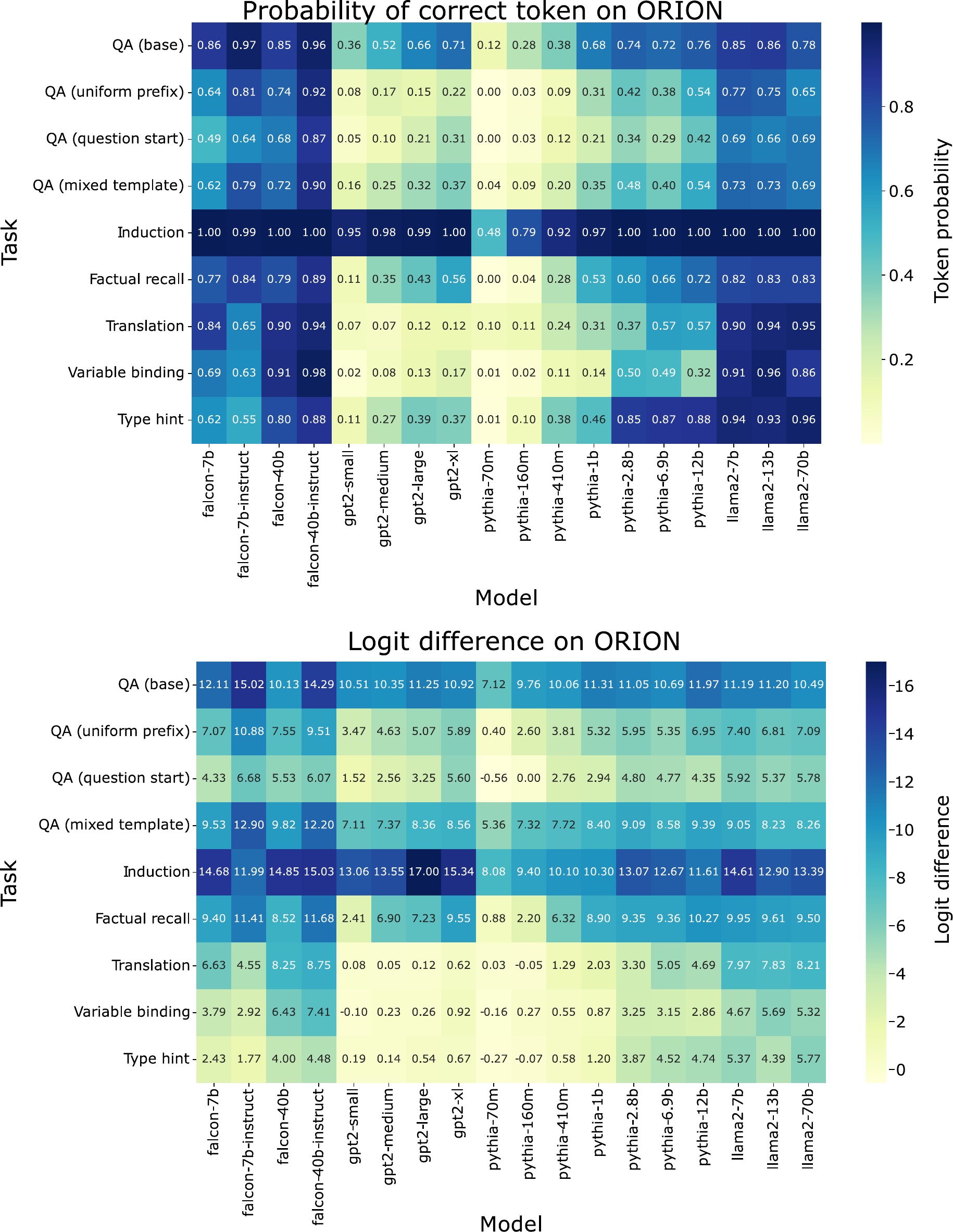}
    \caption{Logit difference and probability of correct token for 18 language models on the tasks from the ORION collection. A logit difference of zero means that the correct logit has on average the same value as the logit corresponding to the answer to a random request in the same context.}
    \label{fig:additionnal_perf}
\end{figure}

We present the performance of the 18 models studied on the ORION collection measured using the logit difference and the probability of the correct token in Figure \ref{fig:additionnal_perf}. The value for the probability of the correct token is used as the normalization factor when computing normalized token probability.

\section{Detailed description of the ORION tasks} \label{app:orion}

\subsection{Discussion about task choice}
For a given problem type, we generate several templates, enabling the creation of several task variations. We use this procedure to generate 15 unique tasks spanning six different abilities. 

We use several criteria in choosing the problem types. First, we choose tasks that have already been studied in the literature to act as reference points for our analysis. This includes factual recall and the induction task. Then, to allow analysis across model scales, we design a simple question-answering task that can be solved by both small and large models. We also create more challenging tasks to explore diverse skills such as coding abilities, tracking the association between an object and its quantity, and tasks involving translation from English to three different languages.

In addition to diversifying the content of the tasks, we create structural modifications to the task template. To that end, we create question-answering templates where the question is before the story in the dataset and templates where the final token of the prompt does not depend on the request. We also create a mixed question-answering task containing prompts from the three variations.\footnote{Given that the mixed-template task is an aggregation of other task variations, we do not include it in the count of 15 unique tasks.}

In this rest of this Appendix, we describe in more detail the process we used to create each task of the ORION collection. We also provide complete example prompts for each task.

\subsection{Question-answering}

\subsubsection{Generating the stories}
We created a set of 100 stories we used in the four variations of the question-answering task. Each story was created by defining:
\begin{itemize}
    \item The name of the main character
    \item The occupation of the main character
    \item The time of day
    \item The season of the year
    \item The city of the story
    \item The action of the story
    \item An order in which the above elements should be introduced in the story
    \item An example story called a “master story”, used as a template to incorporate the new narrative elements
\end{itemize}

The value of each of the narrative elements was uniformly sampled from lists of 3 to 5 different possible values for each field. The lists were generated using manual interaction with ChatGPT.

The 8 narrative elements were combined in a prompt shown in \ref{fig:prompt_gpt4_nanoQA} and completed by GPT-4. The goal of this process was to reduce as much as possible the variations introduced by GPT-4, such that the variables in the generation prompt characterized the generated story as comprehensively as possible.

\subsubsection{Generating the questions}

We manually generated questions and answer prefixes about three different narrative variables: character occupation, city, and name of the main character. For each narrative variable, we created three different phrasings. 

The answer prefixes were either uniform, as shown in Figures \ref{fig:prompt_qa_unif_prefix} and \ref{fig:prompt_qa_question_first} for the task variation with uniform prefix and question at the start, or depended on the variable queried in the question, as shown in Figure \ref{fig:prompt_qa_base} for the base task variation. The base task variation can be solved by smaller models, while only larger models can handle uniform answer prefixes. 

\lstset{
  basicstyle=\small\ttfamily,
  breaklines=true,
  frame=single,
  xleftmargin=0pt,
  xrightmargin=0pt,
  aboveskip=1em,
  belowskip=1em,
  breakindent=0pt,
}

\begin{figure}[h]
    \centering
    \begin{minipage}{0.9\textwidth}
        \begin{lstlisting}
You have to generate a short story that fits in a single paragraph of less than 150 words. It has to respect a list of precise constraints. 

### Narrative elements

The main character is named {character_name}. Their occupation is {character_occupation}. The story takes place in {city}. The time of the day is the {day_time}, and the season is {season}. The time of the day should stay constant in the story. The action performed by the main character is {action}. 

It's crucial that all the elements appear in the story. 

### Order of the narrative elements

The order in which to introduce the narrative elements is imposed. The main priority is to respect the order of apparition I impose. Here is the imposed order in which to introduce the narrative elements. This order is already present in the template story.

{variable_order}

### Template story

You have to generate a story that matches as closely as possible a template story. Your goal is to modify the template story such that all the narrative elements are present, but the general structure (e.g. order in which the narrative element are introduced etc.) is as close as possible to the template story. 

Here is the template story you have to stick to:

"{master_story_text}"

Generate a short story that matches the template story while incorporating the new narrative elements.
        \end{lstlisting}
    \end{minipage}
    \caption{Prompt used to generate the stories for the question-answering tasks. The variables in curly brackets represent placeholders that were replaced by values randomly sampled from manually created lists of possible values.}
    \label{fig:prompt_gpt4_nanoQA}
\end{figure}

\begin{figure}[h]
    \centering
    \begin{minipage}{0.9\textwidth}
        \textbf{Context} 
        \begin{lstlisting}
<|endoftext|>

Here is a short story. Read it carefully and answer the questions below with a keyword from the text. Here is the format of the answer: 'The answer is "xxx".'

The morning sun bathed the streets of Cusco in a warm, golden light, casting long shadows that danced along with the gentle summer breeze. Amidst the bustling city, a tall, slender figure stood on the rooftop of an unfinished building, their eyes surveying the urban landscape below. As the skyline slowly transformed under their careful guidance, it became apparent that the person was no mere observer, but an architect, orchestrating the symphony of steel and concrete. The sound of birdsong filled the air, but it was soon joined by another melody -- the architect's voice, soaring with joy and passion, a song of creation and ambition. And as the last notes faded away, the wind carried a whispered name, the signature of the artist who painted the city with their dreams: Michael.

Answer the questions below.
        
        \end{lstlisting}
        
        \vspace{1em}
        
        \textbf{Request} 
        \begin{lstlisting}
Question: What job does the main character have?

Answer: The answer is "
        \end{lstlisting}
    \end{minipage}
    \caption{Example prompt for the QA (uniform prefix) task.}
    \label{fig:prompt_qa_unif_prefix}
\end{figure}

\begin{figure}[h]
    \centering
    \begin{minipage}{0.9\textwidth}
        \textbf{Request} 
        \begin{lstlisting}
<|endoftext|>

Read the question below, then answer it after reading the story using a keyword from the text. Here is the format of the answer: 'The answer is "xxx".'

Question: What job does the main character have?
                
        \end{lstlisting}
        
        \vspace{1em}
        
        \textbf{Context} 
        \begin{lstlisting}
Story: The morning sun bathed the streets of Cusco in a warm, golden light, casting long shadows that danced along with the gentle summer breeze. Amidst the bustling city, a tall, slender figure stood on the rooftop of an unfinished building, their eyes surveying the urban landscape below. As the skyline slowly transformed under their careful guidance, it became apparent that the person was no mere observer, but an architect, orchestrating the symphony of steel and concrete. The sound of birdsong filled the air, but it was soon joined by another melody -- the architect's voice, soaring with joy and passion, a song of creation and ambition. And as the last notes faded away, the wind carried a whispered name, the signature of the artist who painted the city with their dreams: Michael.

Answer: The answer is "
        \end{lstlisting}
    \end{minipage}
    \caption{Example prompt for the QA (question first) task.}
    \label{fig:prompt_qa_question_first}
\end{figure}

\begin{figure}[h]
    \centering
    \begin{minipage}{0.9\textwidth}
        \textbf{Context} 
        \begin{lstlisting}
<|endoftext|>

Here is a short story. Read it carefully and answer the questions below.

The morning sun bathed the streets of Cusco in a warm, golden light, casting long shadows that danced along with the gentle summer breeze. Amidst the bustling city, a tall, slender figure stood on the rooftop of an unfinished building, their eyes surveying the urban landscape below. As the skyline slowly transformed under their careful guidance, it became apparent that the person was no mere observer, but an architect, orchestrating the symphony of steel and concrete. The sound of birdsong filled the air, but it was soon joined by another melody -- the architect's voice, soaring with joy and passion, a song of creation and ambition. And as the last notes faded away, the wind carried a whispered name, the signature of the artist who painted the city with their dreams: Michael.

Answer the questions below, The answers should be concise and to the point.
                
        \end{lstlisting}
        
        \vspace{1em}
        
        \textbf{Request} 
        \begin{lstlisting}
Question: What job does the main character have?

Answer: The main character is a professional
        \end{lstlisting}
    \end{minipage}
    \caption{Example prompt for the QA (base) task.}
    \label{fig:prompt_qa_base}
\end{figure}

\subsection{Type hint understanding}
Using ChatGPT, we generated three Python code snippets introducing new classes and functions using these classes, as shown in Figure \ref{fig:prompt_type_hint}. The context is a set of variables with a given type. The request asks for a variable name with a particular type. The function definitions do not vary across prompts and are only used to formulate the request.

\lstset{
  basicstyle=\tiny\ttfamily,
  breaklines=true,
  frame=single,
  xleftmargin=0pt,
  xrightmargin=0pt,
  aboveskip=1em,
  belowskip=1em,
  breakindent=0pt,
}

\begin{figure}[h]
    \centering
    \begin{minipage}{0.9\textwidth}
        \textbf{Context} 
\begin{lstlisting}
<|endoftext|>
from typing import List
from math import pi

class Point:
    def __init__(self, x: float, y: float) -> None:
        self.x = x
        self.y = y

class Rectangle:
    def __init__(self, bottom_left: Point, top_right: Point) -> None:
        self.bottom_left = bottom_left
        self.top_right = top_right
        
class Circle:
    def __init__(self, center: Point, radius: float) -> None:
        self.center = center
        self.radius = radius

class Polygon:
    def __init__(self, points: List[Point]) -> None:
        self.points = points

def calculate_area(rectangle: Rectangle) -> float:
    height = rectangle.top_right.y - rectangle.bottom_left.y
    width = rectangle.top_right.x - rectangle.bottom_left.x
    return height * width

def calculate_center(rectangle: Rectangle) -> Point:
    center_x = (rectangle.bottom_left.x + rectangle.top_right.x) / 2
    center_y = (rectangle.bottom_left.y + rectangle.top_right.y) / 2
    return Point(center_x, center_y)

def calculate_distance(point1: Point, point2: Point) -> float:
    return ((point2.x - point1.x) ** 2 + (point2.y - point1.y) ** 2) ** 0.5

def calculate_circumference(circle: Circle) -> float:
    return 2 * pi * circle.radius

def calculate_circle_area(circle: Circle) -> float:
    return pi * (circle.radius ** 2)

def calculate_perimeter(polygon: Polygon) -> float:
    perimeter = 0
    points = polygon.points + [polygon.points[0]]  # Add the first point at the end for a closed shape
    for i in range(len(points) - 1):
        perimeter += calculate_distance(points[i], points[i + 1])
    return perimeter

# Create a polygon
Y = Polygon([Point(0, 0), Point(1, 0), Point(0, 1)])

# Create a rectangle
K = Rectangle(Point(2, 3), Point(6, 5))

# Create a circle
P = Circle(Point(0, 0), 5)
                
        \end{lstlisting}
        
        \vspace{1em}
        
        \textbf{Request} 
        \begin{lstlisting}
# Calculate area
print(calculate_area(
        \end{lstlisting}
    \end{minipage}
    \caption{Example prompt for the type hint understanding task.}
    \label{fig:prompt_type_hint}
\end{figure}

\subsection{Factual recall} \label{app:factual_recall}

Existing open-source datasets created to study factual recall in language models, such as the one introduced in \citep{meng2022locating}, contain relations (e.g. the sport of an athlete) that can only be applied to a subset of the subjects (e.g. only athletes, since asking the sport played by a country does not make sense). This makes it impossible to use causal intervention such as the type required for request-patching (the desiderata “Decomposable” is not fulfilled). Thus, we created two variations of factual recall tasks such that any pair of subject and relation exists. Contrary to the other task from ORION, the retrieval tasks do not involve copying an attribute present in the context, the task requires the model to know the attribute.  

\paragraph{Geography dataset} We used an open-source database\footnote{\url{https://github.com/annexare/Countries}} of countries. We extracted the name, capital city, and continent of each country. 

\paragraph{Geography dataset} Following the process used in \citep{krasheninnikov2023out}, we used a Cross-Verified database (CVDB) of notable people 3500BC-2018AD \citep{laouenan2022cross}. Each individual was ranked by popularity (measured with the “wiki\_readers\_2015\_2018 feature“), and 4000 of the most popular individuals were taken (2000 men and women each). We selected the fields related to the gender, continent, and nationality of each notable person. 

\paragraph{Filtering} For both datasets, we queried the relation about the entity using a few shot setting, as shown in Figure \ref{fig:prompt_factual_recall}. From the raw data extracted from the dataset, we further filtered the list of entities to keep only the ones where GPT-2 was able to answer all the questions related to the entity. The final dataset contains 243 notable people (i.e. 729 questions) and 94 countries (i.e. 282 questions). 

\lstset{
  basicstyle=\small\ttfamily,
  breaklines=true,
  frame=single,
  xleftmargin=0pt,
  xrightmargin=0pt,
  aboveskip=1em,
  belowskip=1em,
  breakindent=0pt,
}

\begin{figure}[h]
    \centering
    \begin{minipage}{0.9\textwidth}
        \textbf{CVDB prompt} 
        \begin{lstlisting}
<|endoftext|>Question: What was the country of Freddie Mercury?
Answer: UK

Question: On which continent did Muhammad Ali live?
Answer: America

Question: What was the country of Fela Kuti?
Answer:
        \end{lstlisting}
        \vspace{1em}
        
        \textbf{Geography prompt} 
        \begin{lstlisting}
<|endoftext|>Question: What is the capital of France?
Answer: Paris

Question: What is the language spoken in Malaysia?
Answer:
        \end{lstlisting}
    \end{minipage}
    \caption{Example prompt for the factual recall task on the CVDB and geography datasets. There is no clear division between context and request in the prompt. In full rigor, the context is composed of a single entity, e.g. '\texttt{Fela Kuti}' in the first prompt, while the request is asking about an attribute, e.g. the country, without filtering as there is a single entity in the context.}
    \label{fig:prompt_factual_recall}
\end{figure}

\subsection{Variable binding}

We were inspired by the shape of grade school math problems from the GSM8K dataset \citep{cobbe2021training}. The goal was to create retrieval tasks that would naturally occur in a chain of thought generated by a model solving a math puzzle. The context contains objects with different quantities. The request asks for the quantity of an object type.

To create the dataset, we picked one sample from the GSM8K dataset and generated variations using ChatGPT. An example prompt can be found in Figure \ref{fig:prompt_variable_binding}.

\begin{figure}[h]
    \centering
    \begin{minipage}{0.9\textwidth}
        \textbf{Context} 
        \begin{lstlisting}
<|endoftext|>John is baking cookies. The recipe calls for 4 cups of flour, 2 cups of sugar, and 6 cups of chocolate chips. How many cups of ingredients in total are needed for the cookies?
        \end{lstlisting}
        
        \vspace{1em}
        
        \textbf{Request} 
        \begin{lstlisting}
We'll add the number of cups of flour (
        \end{lstlisting}
    \end{minipage}
    \caption{Example prompt for the variable binding task.}
    \label{fig:prompt_variable_binding}
\end{figure}

\lstset{
  basicstyle=\scriptsize\ttfamily,
  breaklines=true,
  frame=single,
  xleftmargin=0pt,
  xrightmargin=0pt,
  aboveskip=1em,
  belowskip=1em,
  breakindent=0pt,
  columns=flexible, 
  literate={é}{{\'e}}1
}

\subsection{Translation}

We used ChatGPT-3.5 (referred to as ChatGPT in the main text and the rest of the Appendix) to generate news articles using placeholders instead of real names. We instructed it to add as many names as possible and to prefix each name with a common title, such as "M.".
Then, we asked ChatGPT to translate the text into a non-English language. From the translated text we extracted excerpts that preceded each of the names but did not include any names. These excerpts formed the request. When creating the dataset, the placeholders are replaced by distinct family names from a list generated by ChatGPT.

Using this process, we created three variations with different subjects, target languages, and name prefixes.
\begin{itemize}
    \item Title: "Climate Change: The Unsung Heroes", Prefix: "M.", Target language: French.
    \item Title: "Hidden Wonders Revealed: New Species Discovered in Unexplored Amazon Rainforest", Prefix: "Dr.", Target language: Spanish.
    \item Title: "From Pirates to Naval Heroes: Captains who Shaped Maritime History", Prefix: "Capt.", Target language: German.
\end{itemize}

The entities in the context are the named characters, and their attribute is the sentence in which they appear. The request is asking for a name that appears in a given sentence.

\begin{figure}[h]
    \centering
    \begin{minipage}{0.9\textwidth}
        \textbf{Context} 
        \begin{lstlisting}
<|endoftext|>
Here is a new article in English. Below, you can find a partial translation in French. Please complete the translation.

English:

Title: "Climate Change: The Unsung Heroes"

In an era defined by increasing global temperatures and extreme weather events, the fight against climate change continues on many fronts. While prominent environmentalists and politicians often claim the limelight, behind the scenes, countless unsung heroes have dedicated their lives to combating climate change. This article aims to spotlight the work of these individuals.

At the forefront is M. Jones, a marine biologist who has developed an innovative method for promoting coral reef growth. Given that coral reefs act as carbon sinks, absorbing and storing CO2 from the atmosphere, M. Jones's work has significant implications for climate mitigation. Despite facing numerous hurdles, M. Jones has consistently pushed forward, driven by an unwavering commitment to oceanic health.

Next, we turn to M. Martinez, a climate economist from a small town who has successfully devised a market-based solution to curb industrial carbon emissions. By developing a novel carbon pricing model, M. Martinez has enabled a tangible shift toward greener industrial practices. The model has been adopted in several countries, resulting in significant reductions in CO2 emissions. Yet, despite these successes, M. Martinez's work often flies under the mainstream media radar.

Another unsung hero in the climate change battle is M. Perez, a young agricultural scientist pioneering a line of genetically modified crops that can thrive in drought conditions. With changing rainfall patterns threatening food security worldwide, M. Perez's work is of immense global relevance. However, due to controversy surrounding genetically modified organisms, the contributions of scientists like M. Perez often go unnoticed.

Additionally, the story of M. Thomas is worth mentioning. An urban planner by profession, M. Thomas has been instrumental in designing green cities with a minimal carbon footprint. By integrating renewable energy sources, promoting public transportation, and creating more green spaces, M. Thomas has redefined urban living. While the aesthetics of these cities often capture public attention, the visionary behind them, M. Thomas, remains relatively unknown.

Lastly, we have M. Harris, a grassroots activist working tirelessly to protect and restore the forests in her community. M. Harris has mobilized local communities to halt deforestation and engage in extensive tree-planting initiatives. While large-scale afforestation projects often get global recognition, the efforts of community-level heroes like M. Harris remain largely unsung.

The fight against climate change is not a single battle, but a war waged on multiple fronts. Every victory counts, no matter how small. So, as we continue this struggle, let's not forget to appreciate and honor the unsung heroes like M. Jones, M. Martinez, M. Perez, M. Thomas, and M. Harris who, away from the spotlight, are making a world of difference.
        \end{lstlisting}
        
        \vspace{1em}
        
        \textbf{Request} 
        \begin{lstlisting}
French:
[...]
En intégrant des sources d'énergie renouvelables, en favorisant les transports publics et en créant plus d'espaces verts, M.
        \end{lstlisting}
    \end{minipage}
    \caption{Example prompt for the translation task.}
    \label{fig:prompt_translation}
\end{figure}

\subsection{Induction}

We generated 10 pairs of random strings made from upper and lower-case letters separated by a column. The context contains five enumerations of the pairs. Each enumeration is in a random order. The request is the first half of one of the pairs. An example prompt is shown in Figure \ref{fig:prompt_induction}.

\lstset{
  basicstyle=\tiny\ttfamily,
  breaklines=true,
  frame=single,
  xleftmargin=0pt,
  xrightmargin=0pt,
  aboveskip=1em,
  belowskip=1em,
  breakindent=0pt,
  columns=flexible, 
  literate={é}{{\'e}}1
}

\begin{figure}[h]
    \centering
    \begin{minipage}{0.9\textwidth}
        \textbf{Context} 
\begin{lstlisting}
<|endoftext|>wFCJI:CCwti
axRPX:ISNak
JaVZO:jjVAE
vGuLv:aqCuW
peaXt:uqIWZ
gLbzR:URzLs
XPUgR:QDKMS
IbKIs:YRodj
GpqLd:YRodj
fhVqk:jjVAE
axRPX:ISNak
gLbzR:URzLs
wFCJI:CCwti
GpqLd:YRodj
fhVqk:jjVAE
vGuLv:aqCuW
XPUgR:QDKMS
peaXt:uqIWZ
IbKIs:YRodj
JaVZO:jjVAE
axRPX:ISNak
XPUgR:QDKMS
wFCJI:CCwti
IbKIs:YRodj
gLbzR:URzLs
peaXt:uqIWZ
vGuLv:aqCuW
JaVZO:jjVAE
GpqLd:YRodj
fhVqk:jjVAE
wFCJI:CCwti
GpqLd:YRodj
peaXt:uqIWZ
gLbzR:URzLs
XPUgR:QDKMS
axRPX:ISNak
JaVZO:jjVAE
IbKIs:YRodj
fhVqk:jjVAE
vGuLv:aqCuW
peaXt:uqIWZ
XPUgR:QDKMS
wFCJI:CCwti
JaVZO:jjVAE
IbKIs:YRodj
fhVqk:jjVAE
gLbzR:URzLs
axRPX:ISNak
GpqLd:YRodj
vGuLv:aqCuW
peaXt:uqIWZ
gLbzR:URzLs
GpqLd:YRodj
peaXt:uqIWZ
GpqLd:YRodj
fhVqk:jjVAE
GpqLd:YRodj
XPUgR:QDKMS
peaXt:uqIWZ
        \end{lstlisting}
        
        \vspace{1em}
        
        \textbf{Request} 
        \begin{lstlisting}
wFCJI:
        \end{lstlisting}
    \end{minipage}
    \caption{Example prompt for the induction task.}
    \label{fig:prompt_induction}
\end{figure}

%
%


\end{document}